\documentclass{aa}  
\usepackage{subfigure}

\newcommand{\dgr}{$^{\circ}~$}
\newcommand{\teff}{$\rm T_{eff}$~}

\newcommand{\logg}{$\log{g}$~}

\newcommand{\kms}{\mbox{km s$^{-1}~$}} 
\newcommand{\kmse}{\mbox{km s$^{-1}$}}

\usepackage{threeparttablex}
\usepackage{natbib}

\usepackage{graphicx}
\usepackage{txfonts}
\graphicspath{{./}{figures/}}
\begin{document} 

    \title{Chemo-Dynamical Tagging in the Outskirts:  The Origins of Stellar Substructures in the Magellanic Clouds}

   \titlerunning {Magellanic Periphery Chemistry}
    \authorrunning {Mu\~noz, Monachesi et al.}

    \author{ C\'esar Mu\~noz\inst{1,2}\thanks{E-mail:cesar.alejandro.munoz.g@gmail.com}, Antonela Monachesi\inst{1,2}, David L. Nidever\inst{3}, Steven R. Majewski\inst{4}, Xinlun Cheng\inst{4}, Knut Olsen\inst{5}, Yumi Choi\inst{5,6}, Paul Zivick\inst{7}, Douglas Geisler\inst{1,2,8}, Andres Almeida\inst{4}, Ricardo R. Mu\~noz\inst{9}, Christian  Nitschelm\inst{10}, Alexandre Roman-Lopes\inst{2}, Richard R. Lane\inst{11}, Jos\'e G. Fern\'andez-Trincado\inst{12}
}

\institute{Instituto Multidisciplinario de Investigaci\'on y Postgrado, Universidad de La Serena, Ra\'ul Bitr\'an 1305, La Serena, Chile
 \and Departamento de Astronom\'ia, Universidad de La Serena, Av. Cisternas 1200, La Serena, Chile
 \and Department of Physics, Montana State University, P.O. Box 173840, Bozeman, MT 59717, USA
 \and Department of Astronomy, University of Virginia, Charlottesville, VA 22904-4325, USA
 \and National Optical-Infrared Astronomy Research Laboratory (NOIRLab), 950 North Cherry Avenue, Tucson, AZ 85719, USA
 \and Department of Astronomy, University of California, Berkeley, 501 Campbell Hall 3411, CA 94720-3411, USA
 \and IQVIA
 \and Departamento de Astronom\'ia, Casilla 160-C, Universidad de Concepci\'on, Concepci\'on, Chile.
 \and Departamento de Astronom{\'i}a, Universidad de Chile, Camino El Observatorio 1515, Las Condes, Chile
  \and Centro de Astronom{\'i}a (CITEVA), Universidad de Antofagasta, Avenida Angamos 601, Antofagasta 1270300, Chile
 \and Centro de Investigaci\'on en Astronom\'ia, Universidad Bernardo O'Higgins, Avenida Viel 1497, Santiago, Chile
 \and Instituto de Astronom\'ia, Universidad Cat\'olica del Norte, Av. Angamos 0610, Antofagasta, Chile
 }

   \date{accepted August 21, 2023}

 
  \abstract 
  {
We present the first detailed chemical analysis from APOGEE-2S observations of stars in six regions of recently discovered substructures in the outskirts of the Magellanic Clouds  extending to 20$^{\circ}$ from the LMC  center.  We also present, for the first time, the metallicity and $\alpha$-abundance radial gradients of the LMC and SMC out to 11\dgr and 6\degr, respectively. Our chemical tagging includes 13 species  including  light, $\alpha$, and Fe-peak elements.  We find that the abundances of all of these chemical elements in stars populating two regions in the northern periphery --- along the northern ``stream"-like feature --- show good agreement with the chemical patterns of the LMC, and thus likely have an LMC origin. For substructures located in the southern periphery of the LMC, we find more complex chemical and kinematical signatures, indicative of a mix of LMC-like and SMC-like populations. However, the shouthern region closest to the LMC shows better agreement with the LMC whereas that closest to the SMC shows a much better agreement with the SMC chemical pattern.  When combining this information with 3-D kinematical information for these stars, we conclude that the southern region closest to the LMC has likely an LMC origin whereas that closest to the SMC has an SMC origin and the other two southern regions have a mix of LMC and SMC origins.   Our results add to the evidence that the southern substructures of the LMC periphery are the product of close interactions between the LMC and SMC, and thus likely hold important clues that can constrain models of their detailed dynamical histories. 

}


   \maketitle
%

\section{Introduction}
\label{sec:intro}

According to the standard cosmological paradigm of cosmological structure formation, halos grow in mass hierarchically via the accretion and merger of smaller halos \citep[e.g.,][]{WhiteRees1978}. Galaxies form inside the more massive halos, where the gas can collapse and form stars. The self-similarity of the $\Lambda$CDM paradigm implies that accretion and mergers are expected to also take place in dwarf galaxies \citep[e.g.,][]{Diemand2007}, which in turn are also the most common type of galaxies in the Universe. However this prediction is poorly constrained by observations,
mostly because dwarf galaxies are intrinsically faint.

The Large and Small Magellanic Clouds (LMC and SMC, hereafter) are our closest pair of interacting dwarf galaxies, at distances of $\sim$50 and 64 kpc, respectively \citep{Pietrzynski2019,Graczyk2020}. They are, thus, unique and important systems to investigate in great detail not only their formation but also their interaction history and evolution, and to put constraints on dwarf galaxy formation models. Consequently, they have been the main target of several dedicated surveys (see below). 

It is now known that the Magellanic Clouds' (MC's) stellar populations extend much farther than previously thought \citep[e.g.,][]{Nidever2011,Belokurov2016}, and that they suffered close interactions that likely produced both the gaseous Magellanic Bridge \citep{Hindman1963, Harris2007}
and a likely large amount of still undetected stellar debris \citep[][although see \citealt{Luri2021} for detection of stars in the Magellanic Bridge]{Besla2012, Lucchini2021}. 
In addition, very precise proper motion (PM) measurements confirmed that the Magellanic Clouds (MCs) are on their first infall towards the Milky Way (MW) \citep{Kallivayalil2013,Besla2007}. This scenario was supported by several observations, but in particular by the recent observational evidence of a local MW dynamical wake, which was predicted to trail the orbit of the LMC, and a large-scale MW overdensity, also predicted to exist across a large area of the northern Galactic hemisphere \citep{gomez2016,Nicolas2019,Garavito2021,Conroy2021,Erkal2021,Peterson2021}.


To understand the hierarchical assembly involving such dwarf galaxies, it is important to study their periphery, which is most affected by tidal interactions and where the fossil record of interaction and past accretion events is most long-lived.
During the last decades large surveys providing both homogeneous, wide-field photometry as well as deep individual fields (e.g., DES: \citealt{Bechtol2015}, SMASH: \citealt{Nidever2017}, MagLiteS: \citealt{Drlica-Wagner2016}, MagES: \citealt{Cullinane2020}, DELVE: \citealt{Drlica-Wagner2021}) have been dedicated to imaging the outskirts of the MCs. These observations have unveiled the  highly disturbed LMC/SMC disks and other faint stellar structures in unprecedented detail, including a northern stream-like feature \citep{Mackey2016, Mackey2018}, ring-like structures \citep{Choi2018a, Choi2018b}, and a diffuse, extended stellar component \citep{Nidever2019}. 

In addition to dedicated MC surveys, thanks to the {\it Gaia} space telescope \citep{gaiacollaboration16}, we can now also obtain a homogeneous, although not as deep, view of the MC outskirts, by selecting relevant stars using proper motion criteria. 
\citet{Belokurov2019} used red giant branch (RGB) stars from {\it Gaia} DR2 and discovered remarkable, extended streams of stars in the periphery of the Clouds that highlight the complexity of this interacting system. These features could only recently be detected so sharply due to the contiguous, uniform coverage of {\it Gaia}, which is crucial to observe stellar streams that  possess very low surface brightness (${\sim}32$ mag/arcsec$^2$) and cover a large extent on the sky (${\sim}10$s of degrees). While the discovery of these structures around the LMC provides important insights into the MC system, their exact origin remains unclear.  It is possible that both a tidally disturbed disk as well as a classical accreted halo contribute to the diffuse substructures in the MC's periphery. 

To help disentangle the nature of these features, we need both stellar kinematical and chemical abundance information of the substructures. 
The kinematics of these substructures were recently  investigated in \citet[][hereafter C22]{Cheng2022}, from new Apache Point Observatory Galactic Evolution Experiment-2S (APOGEE-2S, \citealt{Majewski2017}, Majewski et al. in prep.) observations of six fields around the MCs periphery. The fields where placed on top of the overdensities of stars that have been detected, reaching out to 20$^{\circ}$ on the north and 15$^{\circ}$ on the south from the LMC center. C22 performed a detailed kinematical study by combining the APOGEE-2S radial velocities and the proper motions provided by {\it Gaia} DR3 data \citep{GaiaCollaboration2021} and found that stars in the southern region have extreme space velocities that are distinct from, and not a simple extension of the LMC disk. On the other hand, the stars in the northern substructure are consistent with being part of the LMC disk (see also \citealt{Cullinane2021}, who explored the kinematics of the northern arc in the LMC periphery). 
It was also found in C22 that the combination of LMC and SMC debris produced from their interaction is a plausible explanation for the extreme velocities in the southern periphery of the LMC, although it is not possible to rule out other origins.

To further constrain the origin of these overdense regions, information about their chemical abundances is needed, which is also available from the APOGEE-2S observations.  Chemical abundances of the Magellanic system contain relevant information about star formation, evolution and their interaction with the interstellar medium. The detailed chemical characterization of a number of elements available from APOGEE-2S will allow us to understand the relation between the six regions and the LMC, SMC and the MW better. Also, we can investigate different starbursts 
produced in the Magellanic system and how the interstellar medium (ISM) polluters such as  Asymptotic Giant Branch (AGB) stars and supernovae among others, influenced the chemical fingerprint observed nowadays in the Magellanic system and in their periphery. Moreover, we will be able to offer more detailed information about the age-metallicity relation in the LMC \citep{Piatti2013,carrera2008}. 

In this paper, we follow-up and complement the work  C22 by presenting a detailed analysis of the chemical abundance patterns of the six fields in the MCs periphery studied by C22.
We compare our findings with the abundances of the LMC and SMC studied by \citet{Nidever2020} as well as with the abundances of other MW dwarf galaxies \citep{Hasselquist2021} in a consistent and homogeneous way, by using the APOGEE DR17 database \citep{Abdurro2022} for all the sources analyzed.
 

With this work, we aim to constrain further constrain the origin of the overdensity regions around the MC periphery and investigate any
possible association of the stars in the LMC periphery to those in either the LMC or SMC.

The outline of this paper is as follows. In Section \ref{sec:data}, we describe the APOGEE-2S observations and the selection of the sample of stars that we analyze in this work, as well as the sample of MW  stars used for comparison. Section \ref{sec:results} presents our general results in terms of the metallicity and $\alpha$-abundance radial profiles and metallicity distribution function. In Section \ref{sec:discussion}, we discuss in detail  abundances for a variety of elements  and interpret the results for each one of the six regions analyzed, separately.  Finally, we summarize and conclude our work in Section \ref{sec:summary}.

\section{Data}
\label{sec:data}


\begin{figure}
\centering
\includegraphics[width=3.5in,height=3.5in]{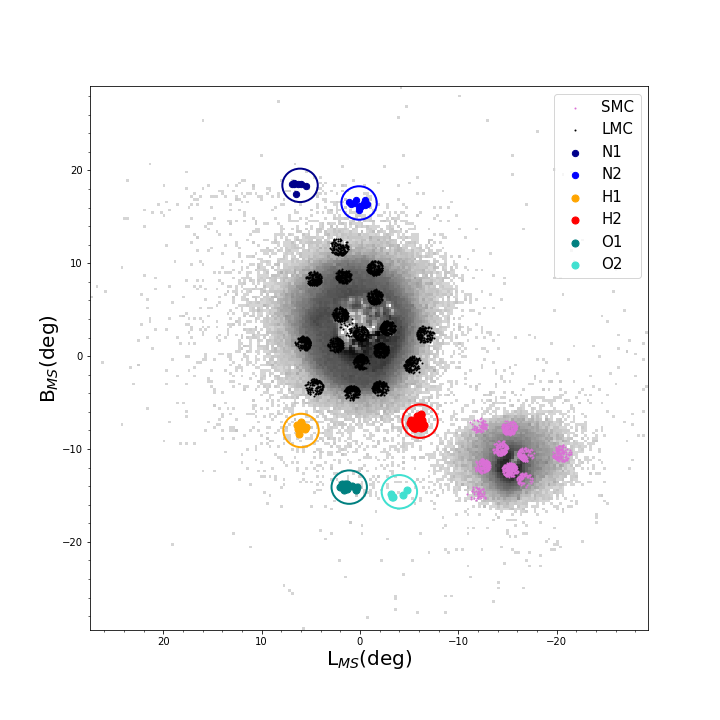}

\caption{Distribution of the six fields in the substructures as well as LMC and SMC fields observed by APOGEE-2  in Magellanic Stream coordinates ($L_{MS}, B_{MS}$).}
\label{MapCoord}
\end{figure} 

Our study is focused on the chemical analysis of six APOGEE fields placed on substructures identified by \citet{Belokurov2019} in the outskirts of the LMC and SMC (see Figure \ref{MapCoord}). 
APOGEE \citep{Majewski2017} was originally part of Sloan Digital Sky Survey III \citep[SDSS-III][]{Eisenstein2011} as a high resolution, near-infrared spectroscopic survey of Galactic stars accessible from the Sloan 2.5-m Telescope 
\citep{Gunn2006}, but the survey was expanded as APOGEE-2 (Majewski et al., in prep.) in SDSS-IV \citep{Blanton2017} to include measurements made with the du
Pont 2.5-m telescope \citep{Bowen1973} in the
Southern Hemisphere, partly motivated by the desire to probe the Magellanic Clouds. 
Targeting for the APOGEE surveys, including for the Magellanic Cloud program described here, is described
in \citet{Zasowski2013,Zasowski2017},  \citet{Beaton2021} and \citet{Santana2021}.
We analyze chemical abundances measured from spectra that were obtained from the second high resolution, near-IR, APOGEE spectrograph built for APOGEE-2 \citep{Wilson2019}.

Our observations were obtained through the Chilean National Telescope Allocation Committee (CNTAC) program CN2019A-30 (PI: A.\ Monachesi). Table \ref{tab:coordinate} lists the location of the six APOGEE fields, named following C22 labels, which cover a diameter of 2\dgr each, as well as their field name according to APOGEE-2S and total visits and $H$ magnitude depth. These fields reach out to 20$^{\circ}$ from the LMC center, which extend the MCs program from the APOGEE survey by about 10$^{\circ}$.
Each APOGEE plate has 300 fibers, therefore, $\sim$260 science targets and $\sim$40 calibration targets were observed per field. At the location of our fields (see Table \ref{tab:coordinate}), the LMC surface brightness is $\sim$30 mag/arcsec$^2$, thus only a few stars on each plate are from the MC system --- the vast majority of stars are MW field star contaminants (or ``filler'' targets; see APOGEE-2 targeting references cited earlier), which we need to remove from our sample.

The data were reduced with the standard APOGEE reduction pipeline \citep{Nidever2015}, which has been updated for improved calculation of radial velocities, especially for fainter stars like ones we stars such as  analyze here.  The chemical abundances
used here were obtained from  the APOGEE Stellar Parameters and Chemical Abundance Pipeline \citep[ASPCAP,][]{Garcia_Perez2016} as given in SDSS Data Release 17 \citep[DR17,][]{Abdurro2022}.
ASPCAP is based on the FERRE2 code of \citet{Prieto2006} and uses a grid of MARCS stellar atmospheres \citep{Gustafsson2008,Jonsson2020}  with an $H$-band line list from \citet{Smith2021}, an update of the previous version by
\citet{Shetrone2015}. These atmospheres and the line
list are combined to create a grid of synthetic spectra \citep{Zamora2015}
using the Synspec code \citep{Hubeny2011}.  Some elements --- in particular,  Ca and Mg used here --- required the use of non-LTE calculations \citep{Osorio2020}. Once created, the library of synthetic spectra
are used to find a best match to each observed spectra to determine stellar
parameters and chemical abundances.


The APOGEE spectra provide not only chemical abundances but also precision (to a few hundred m s$^{-1}$)
heliocentric radial velocities. These, together with the proper motions provided by {\it Gaia}, can be used to obtain the internal 3-D velocities of the stars with respect to the LMC or SMC reference frame. This is relevant to understanding the origin of the stars in the substructures. 

C22 calculated the 3-D velocities of the APOGEE stars in the six substructures, as well as in the central LMC region,  with respect to the LMC, using the model presented in Olsen et al.\ (in prep.; see also \citealt{Choi2022} for a description of this model). In brief, the model follows the formalism by \citet[][hereafter vdM02]{vanderMarel2002}, 
which describes the relationship between the proper-motion vector
and the orthogonal velocity components in the plane
of the sky (as defined in Equation (1) of vdM02). The kinematical model was obtained after fitting 12 parameters jointly with the heliocentric velocities and {\it Gaia} proper motions of $\sim$15,000 AGB and RGB stars. The best-fit parameters of the model are found in Table 1 of \citet{Choi2022}. Then, using the orientation of the LMC disk obtained from the model and the transformations of the 3-D motions to a cylindrical coordinate system (from the equations presented in vdM02), C22 derived $V_r$ and $V_{\phi}$, the radial and rotational motions projected onto the
LMC disk plane, and the vertical velocity $V_z$, the motion perpendicular to the disk plane (where a positive $V_z$ is toward the Sun) of all the stars in the substructures.

We also repeat this derivation for the substructure stars with respect to the SMC instead of the LMC. To do this, we use the formalism and kinematic model for the SMC presented in \citet[][hereafter Z21]{Zivick2021}, which follows a similar process as C22's adaption of vdM02's formalism. However, there are two key differences between the LMC and SMC in the formalism. 1) The model of the  LMC assumes a thin disk for all stars present, which allows for the calculation of the distance of individual stars in the frame of the LMC (a requirement for deriving internal velocities).
In the case of the SMC, to calculate  the internal velocities,  we explore three distance assumptions: 50 kpc, 60 kpc, and 70 kpc for the stars. For simplicity, we choose to calculate the velocities as if all substructure stars sit at 60 kpc, roughly consistent with the distance to the SMC center as determined by RR Lyrae stars (Jacyszyn-Dobrzeniecka et al. 2017). 2) The LMC model assumes a state of equilibrium, which is not the case for the SMC. As shown in Z21, an additional velocity component is required to describe the relative motions of SMC stars, namely the tidal expansion component due to the ongoing tidal disruption occurring in the SMC due to interactions with the LMC. Using the aforementioned assumed distance of 60 kpc and the Z21 formalism and best-fit kinematic model parameters, we derive the internal velocities, $V_r$, $V_{\phi}$, $V_z$, with respect to the SMC for all stars in the substructures as well as in the central SMC region.


In this work, we use the 3-D velocity information jointly with the chemical abundances of the stars in the substructures to understand their origin.

\begin{table*}

\centering

\caption{Regions, number of  stars, coordinates, H magnitude, number of visits  and angular distance from the LMC center for the six substructures analyzed in this article.}
\label{tab:coordinate} 
{
\begin{tabular}{ l  c c c c  c  c c     }
\hline 
\hline
{\small{}Region} & N$_{\rm Stars}$  & Field & RA  &DEC  & H & N$_{\rm  Visits}$  &Angular Distance  \tabularnewline

   &  &     & \small{(h:m:s)}  &  \small ($\,^{\circ}{\rm }$:$^{\prime}$:$^{\prime\prime}$)& (mag)&    & (\degr)    \tabularnewline

\hline 
 N1  & 7  & 261-27-C &	06:21:43  &	$-$53:34:00 & 13.36 -- 14.08  &  9  & 17.26 \\ 	
 N2  & 13 & 264-33-C &  05:30:29  &	$-$55:42:00 & 12.88 -- 13.88  &  9  & 14.00 \\ 	
 H1  & 7  & 291-25-C &  07:24:17  &	$-$79:03:00 & 12.70 -- 13.72  & 10  & 11.70 \\ 	
 H2  & 27 & 293-37-C &	03:23:46  &	$-$77:11:00 & 12.65 -- 13.88  &  9  & 11.30 \\ 
 O1  & 10 & 299-28-C &  05:35:19  &	$-$86:14:00 & 12.50 -- 13.20  & 10  & 16.51 \\ 	
 O2  & 5  & 301-33-C &	01:48:10  & $-$84:08:00 & 11.89 -- 13.89  & 11  & 17.87 \\

\hline 

\end{tabular}
}
\end{table*}

\subsection{Selection of MC stars}

With the primary goal of studying the chemical abundances of the stars in the MC substructures, which are more challenging to measure than radial velocities and bulk metallicities, we first take the sample of MC stars from the selection made by C22, as briefly described below.

\begin{figure}
  \centering
  \begin{minipage}{0.85\linewidth}
    \includegraphics[width=\linewidth]{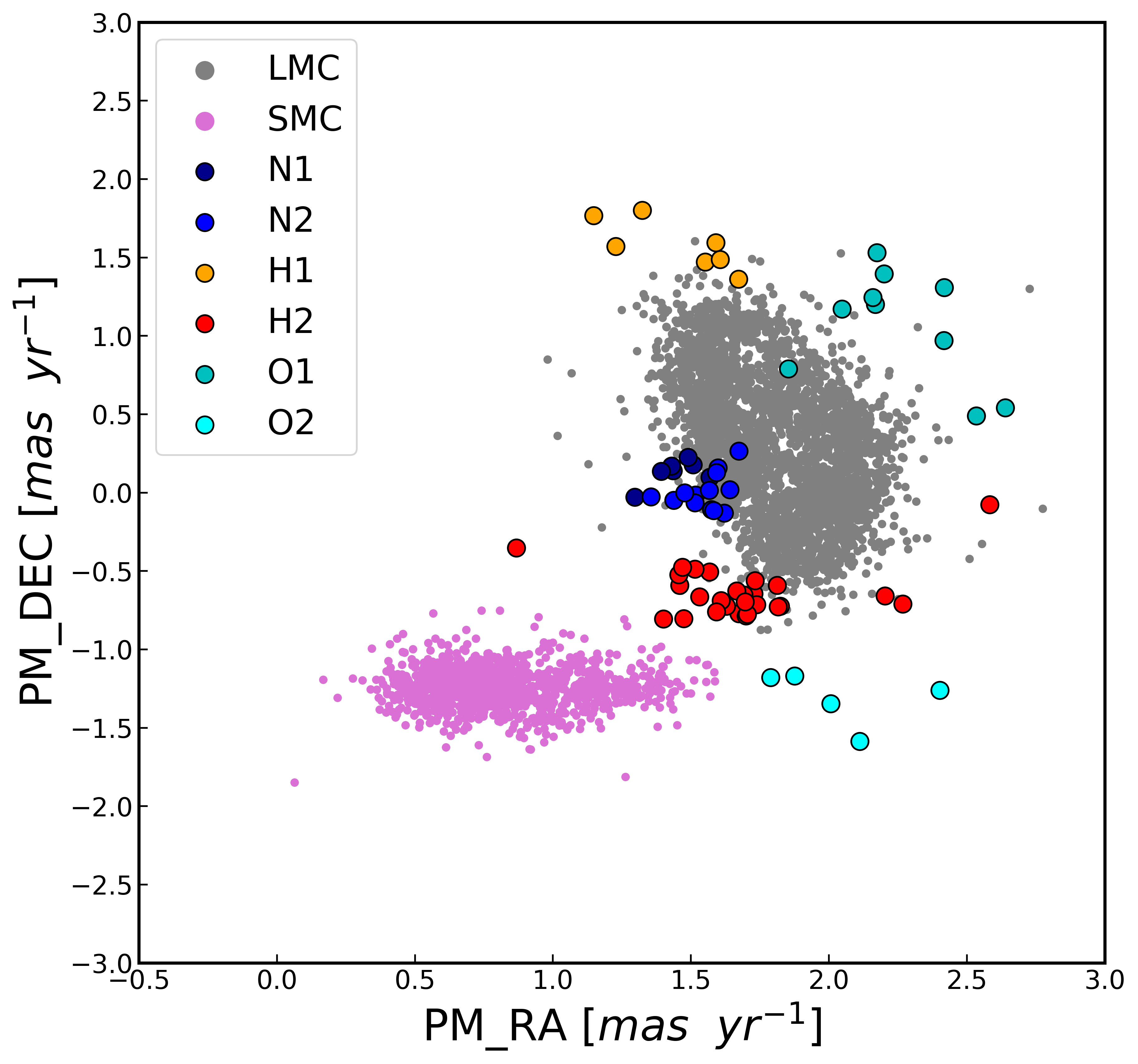}
    \label{fig:pic1}
  \end{minipage}
  \hfill
  \begin{minipage}{0.85\linewidth}
    \includegraphics[width=\linewidth]{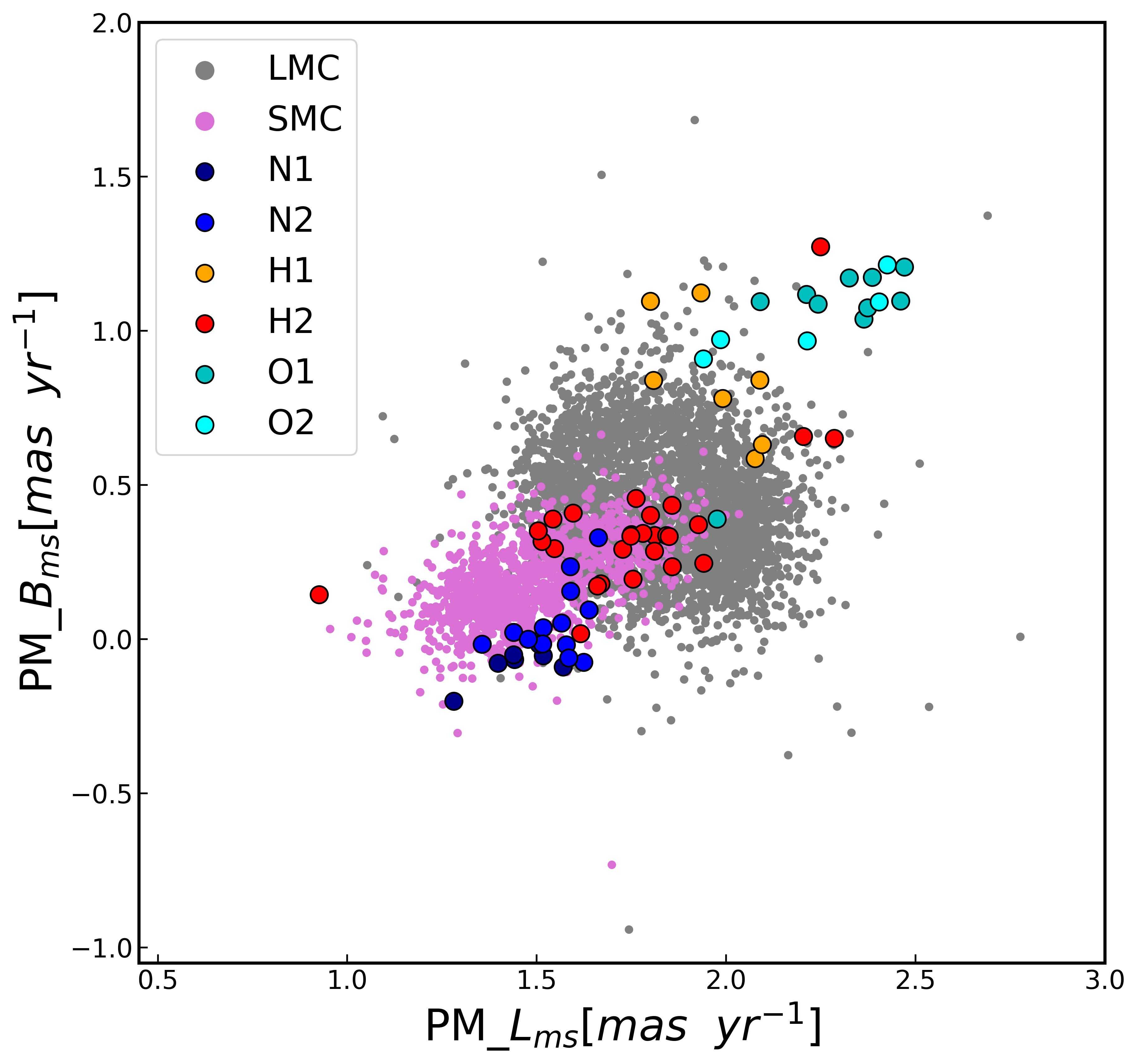}
    \label{fig:pic2}
  \end{minipage}
  \caption{Figure in the top is {\it Gaia} EDR3 proper motions for stars in the six regions analyzed as well as LMC and SMC stars. The figure in the bottom is {\it Gaia} EDR3 Proper Motion in Magellanic Stream coordinates ($L_{ms},B_{ms}$).}
  \label{Proper_Motion}
\end{figure}

The sample used by C22 was selected using parameters delivered by {\it Gaia} EDR3 
and presented in APOGEE DR17. 
The specific selection criteria are: stars with $G$ $<$ 17.5 within 30$^{\circ}$ of the center
of the Magellanic Stream coordinate system \citep{Nidever2008} and that have similar proper motions to the LMC. This PM selection is shown in Figure~\ref{Proper_Motion} where the stars in our final list in each substructure are color coded. We can see in that figure that some of the stars in these substructures exhibit a  slightly  larger proper motion than that of the  LMC  and/or SMC. Since these stars are located in the outskirts of the MCs, this might be due to differences in velocities/distances from the main bodies. Nevertheless, the position of  the stars in proper motion are also in agreement with the selection range  made by \citet{Cullinane22} and can be associated with the MCs system. 

\begin{figure}[t]
\centering
\includegraphics[width=3.2in,height=2.3in]{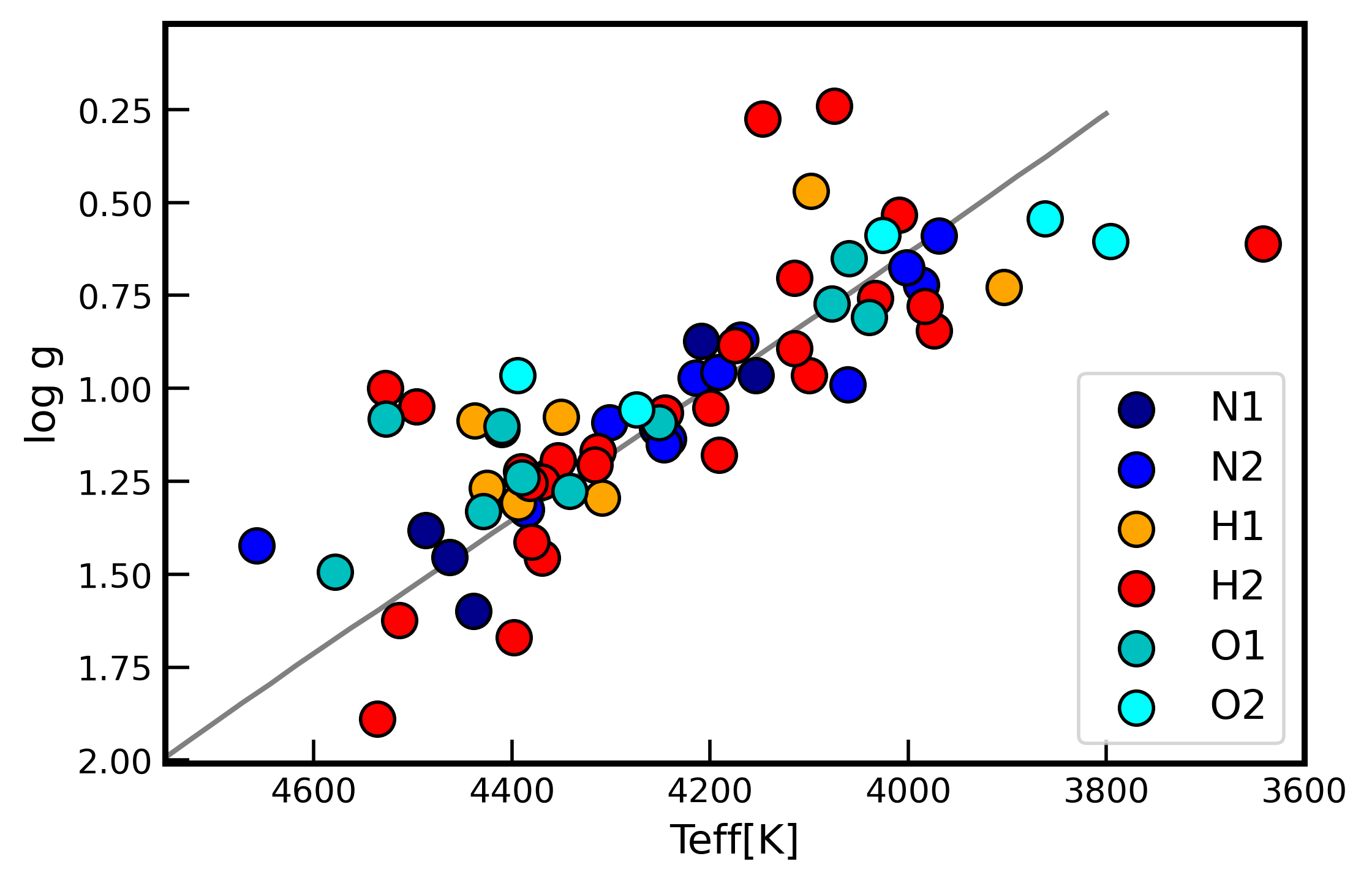}
\caption{The \logg versus\ \teff distribution from APOGEE-2 data release (DR17) for the six regions listed in Table \ref{tab:param1}. The solid
grey line represents a Dartmouth RGB isochrone  with a metallicity of -1.0 dex and age of 11 Gyr \citep{Dotter2008}.}
\label{TeffvsLogg}
\end{figure}

Additional selection criteria were used by C22 in stellar parameters to  constrain the sample to stars along the RGB of the LMC. The specific selection was: \teff $<$ 5400 K, \logg $<$ 4.0  in addition to the magnitude cut mentioned above.
To avoid foreground stars from the MW, stars with parallax $\pi > 0.2$ mas or Galactic latitude $|b| < 5^{\circ}$ were removed. Also, only stars within the proper motion space of the MCs were considered (see Figure \ref{Proper_Motion}), as well as stars with 100 $<$ $V_{\rm Helio} < $ 350 km s$^{-1}$, which is the typical heliocentric velocity range for MC stars \citep[see][]{Nidever2020}. 
Our final sample was very restricted in both \teff and \logg (see Figure \ref{TeffvsLogg}).

\begin{figure*}
\centering
\includegraphics[width=7.0in,height=5in]{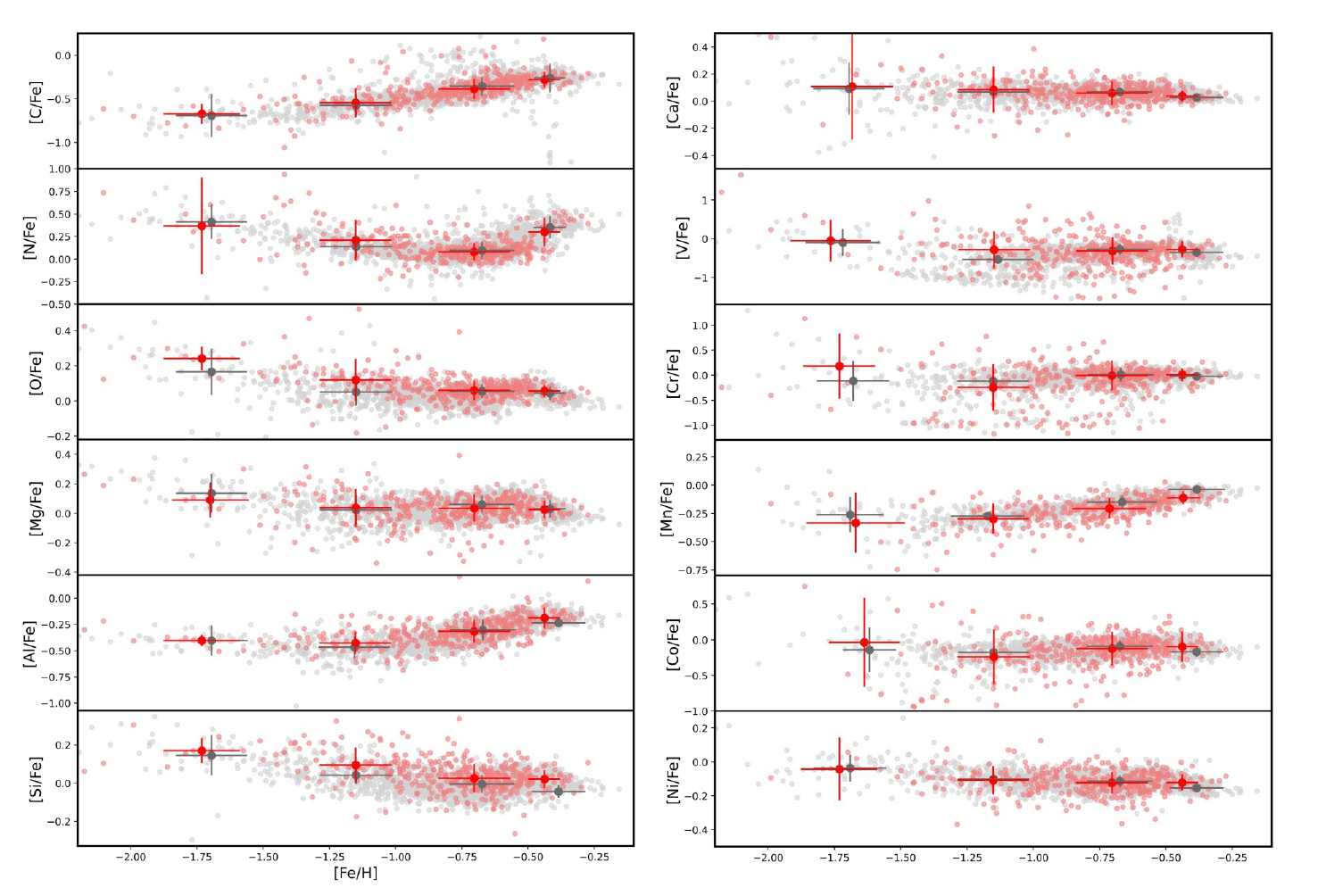}
\caption{Comparison of the APOGEE-2 LMC stars with different S/N. The small red dots are stars with S/N between 35 and 70 and the small gray dots represent stars with S/N$>$70.  
We divided the sample into four by metallicity over the range $-$2.0 to 0.0 dex to obtain the median and dispersion of the stars at different representative metallicities.
The filled large red and gray circles with error bars represent these values, respectively.}
\label{SNR_comp}
\end{figure*}

Additionally, in this study we perform an extra constraint, not used in C22, related to the signal-to-noise ratio (S/N) of the sample, which is critical to the chemical analysis. Thus, we use in this work a sub-sample of C22's sample of stars.
Since our sample is very small, containing few MC-selected stars, we need a S/N cut that is high enough to perform a chemical analysis but at the same time not too restrictive so that we retain as many stars as possible. \citet{Nidever2020} compared the results obtained for metallicity ([Fe/H]) and $\alpha$-elements ([Ca/Fe], [Si/Fe], [Mg/Fe]) when using a sample of stars with low S/N ($\sim$40) and with a sample of stars with higher S/N $>$70. They found that the sample with low S/N was good enough to analyze the chemical patterns of the LMC and SMC, at least for the elements mentioned.  Following their approach, we performed a quality check of our data for the elements we will be using in this work, taking into account a minimum S/N of 35. In Figure \ref{SNR_comp}, we compare APOGEE-2 DR17 data for LMC stars selected as members by \citet{Nidever2020} and for all the elements analyzed in this study. In this figure, we compare each chemical element versus metallicity for the LMC stars with a S/N greater than 70 (gray in Figure \ref{SNR_comp}) and for the stars with an S/N between 35 to 75 (red dots in Figure \ref{SNR_comp}). We divided the metallicity range into four main regions 
and plotted the results for each of the 13 elements analyzed. In each of these four regions, we calculated the median and standard deviation of each group (large filled circle with error bars).  We find that for the $\alpha$ and light elements the medians are in very good agreement when taking into account the dispersion, especially considering that this scatter is greater for the group with more metal-poor stars and with lower S/N. In the case of Fe-peak elements, this difference is more noticeable, and, therefore, we decided to use only stars with S/N greater than 60 in our analysis of these elements. These results show that a S/N cut of 35 is good enough for our data. It is worth mentioning that while the minimum S/N of our data is 35, the maximum is 192 and 44\% of the selected stars listed in Table \ref{tab:param1} have an excellent S/N, greater than 70. 

Our final sample contains 69 stars, in contrast to the 84 stars used in the \citet{Cheng2022} analysis. The number of selected stars in each region is listed in Table \ref{tab:coordinate}. 

\subsection{Milky Way foreground stars}
\label{subsec:mwforeground}

We select a sample of foreground MW stars, which we use throughout our analysis as a comparison sample. This sample also contains stars from APOGEE DR17, but from two datasets: First we take all stars at distances greater than 10\dgr up to 18\dgr from the center of the LMC (which includes our six regions in the LMC/SMC outskirts), but that do not meet the selection criteria as members of LMC/SMC or the six substructures analyzed here. These stars 
have observed radial velocity and proper motions consistent with being MW stars. We also include the sample of halo MW stars from \citet{Hayes2018}. These stars were selected by \citet{Hayes2018} as halo population members using ASPC parameters with selection criteria described therein.

\begin{table*}
\centering
\caption{Basic parameters for the six substructures described in the text.}
\label{tab:param1} 
{

\begin{tabular}{ccccccccc}
\hline 
\hline
{\small{}Region} & N$_{\rm Stars}$ & $\langle$S/N$\rangle$ &S/N$_{\rm Min}$ & S/N$_{\rm Max}$ &$\langle${\small{}RV$_{\rm H}$}$\rangle$  & $\sigma$(\small{RV$_{\rm H}$}) &  $\langle$\small{[Fe/H]}$\rangle$ & $\sigma$(\small{[Fe/H]}) \tabularnewline

   &  &     &  &  &\kmse &   \kmse &    & dex\tabularnewline

\hline 
N1   & 7 & 68 &  37 & 113  & 316.2 &13.5  & $-$1.03 &  0.19 \\
N2   & 13 &  80 &  48 & 129 & 293.9 & 10.4 & $-$1.01  & 0.22 \\
H1  & 7 &75 &  38& 156 & 261.1 & 19.5 & $-$1.25  &  0.34  \\
H2  & 27  &66 &  35& 147 & 168.6 & 26.5 & $-$1.00  &  0.22  \\
O1   & 10  & 65 &  39& 121 & 198.0 & 19.2 & $-$1.03  &  0.26  \\
O2   & 5   & 153 &  35& 192 & 168.6 & 9.0 & $-$1.14  &  0.22  \\

\hline 

\end{tabular}
}
\end{table*}

\begin{figure*}
\centering
\includegraphics[width=0.85\hsize,angle=0]{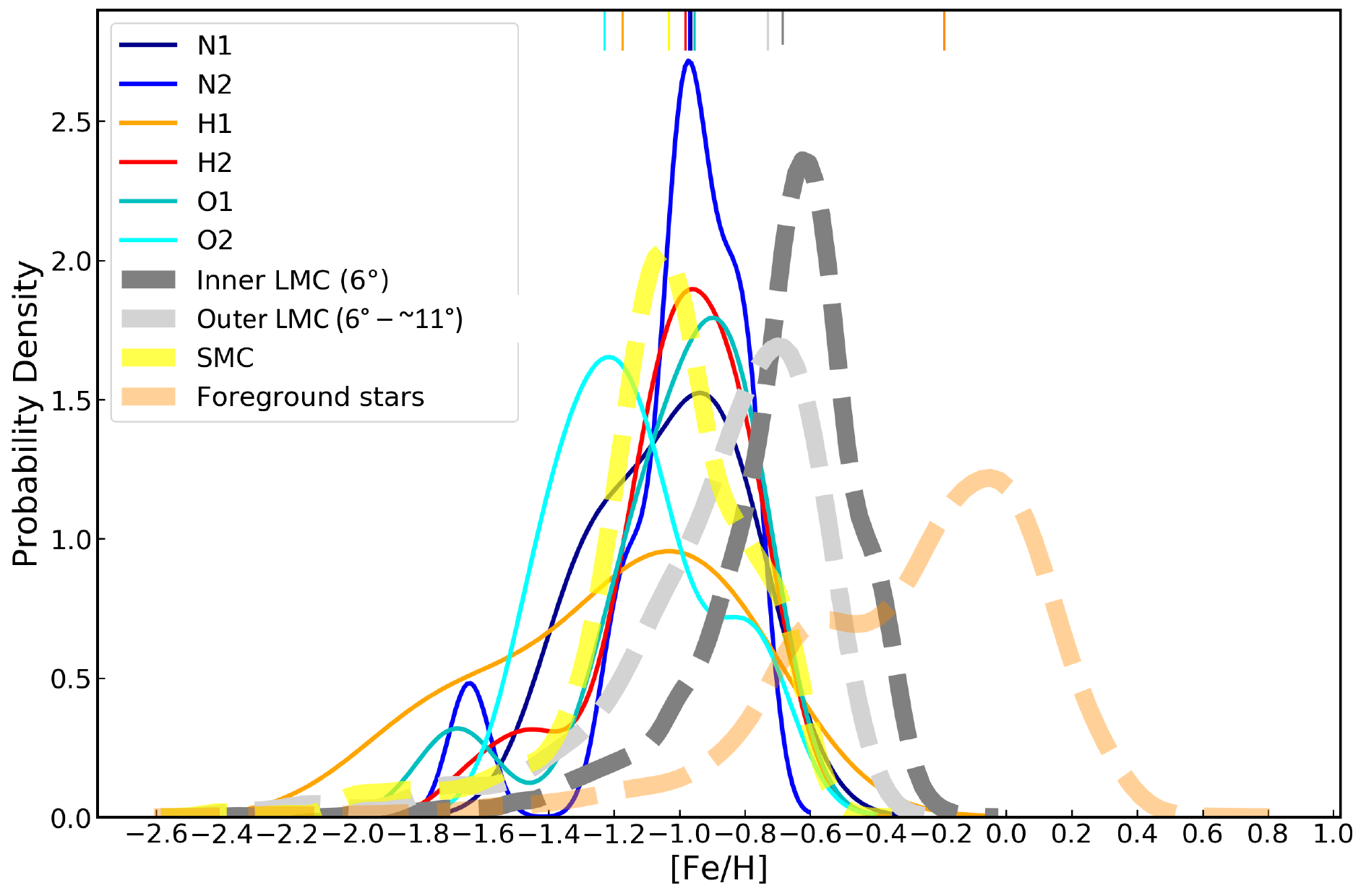}
\caption{Normalized probability density functions applied to the observed metallicities in each region listed in Table \ref{tab:param1}, and including  APOGEE measurements of more interior parts of the LMC and the SMC. The median metallicity for each region is indicated by a line 
at the top of the figure.}
\label{MetDensity}
\end{figure*}

\begin{figure*}
\centering
\includegraphics[width=2\columnwidth,angle=0]{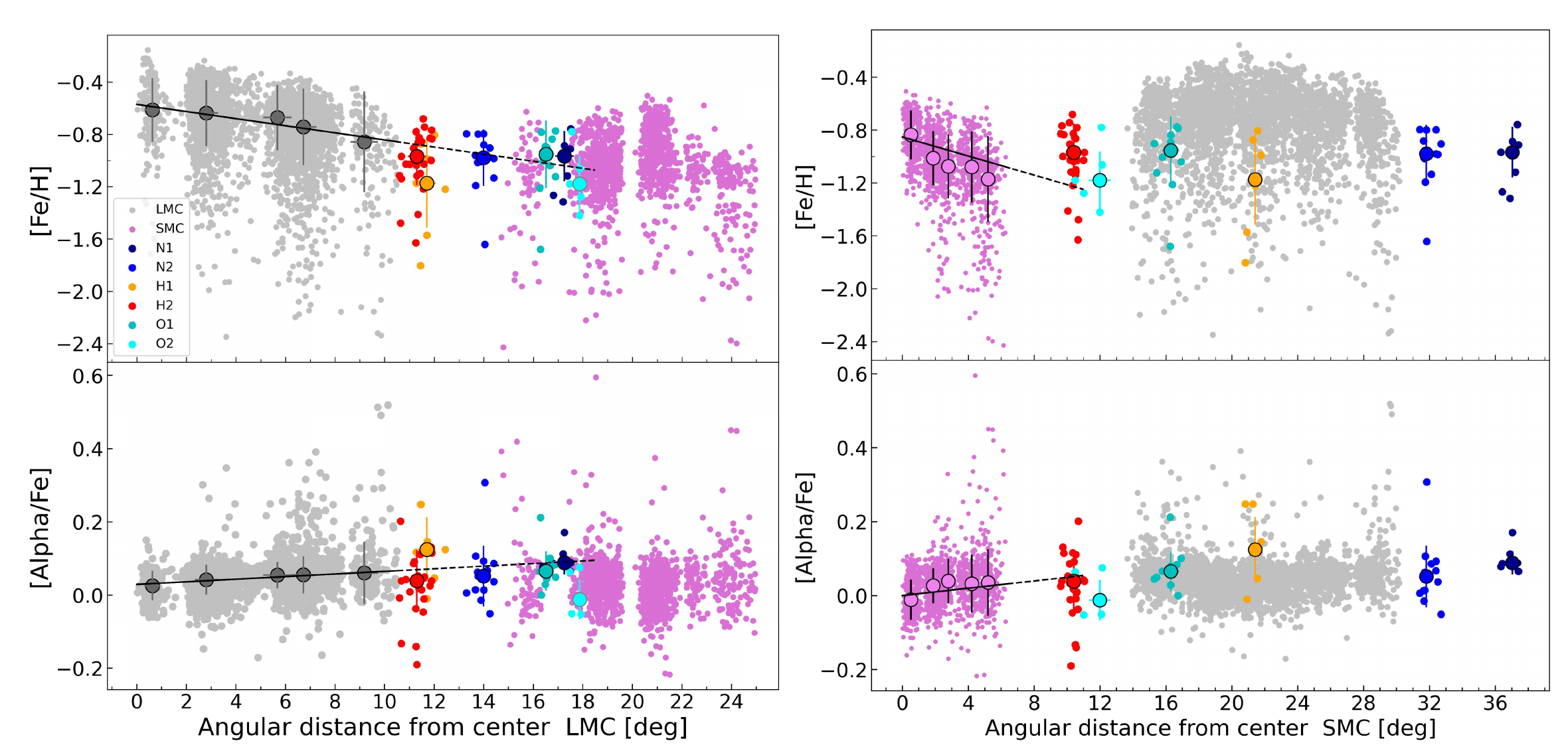}

\caption{
Metallicity and $\alpha$-abundance, 
defined here by [(Mg+Si+Ca)/Fe],
radial profiles of our sample from the LMC  center (left panel) and SMC center (right panel). {\it Left panels:} The filled large symbols with error bars represent the median of each region (listed in Table \ref{tab:param1}) and the bars represent the standard deviation of each region. The solid lines are linear fits made using the APOGEE-2 LMC data inside 11\dgr that we extrapolated (dashed lines) to show the predicted values from the LMC to its outskirts, where our six APOGEE fields lie (shown also with color-coded points). {\it Right panels:}  The filled large symbols with error bars represent the median values for each region (listed in Table \ref{tab:param1}) and the error bars represent the standard deviation of each region.  The solid lines are linear fits made using the APOGEE SMC data inside 6\dgr that we extrapolated to show the predicted values from the SMC out to where our six APOGEE fields.
}
\label{prof_LMC_SMC}
\end{figure*} 

\begin{figure*}
\centering
\includegraphics[width=0.75\hsize,angle=0]{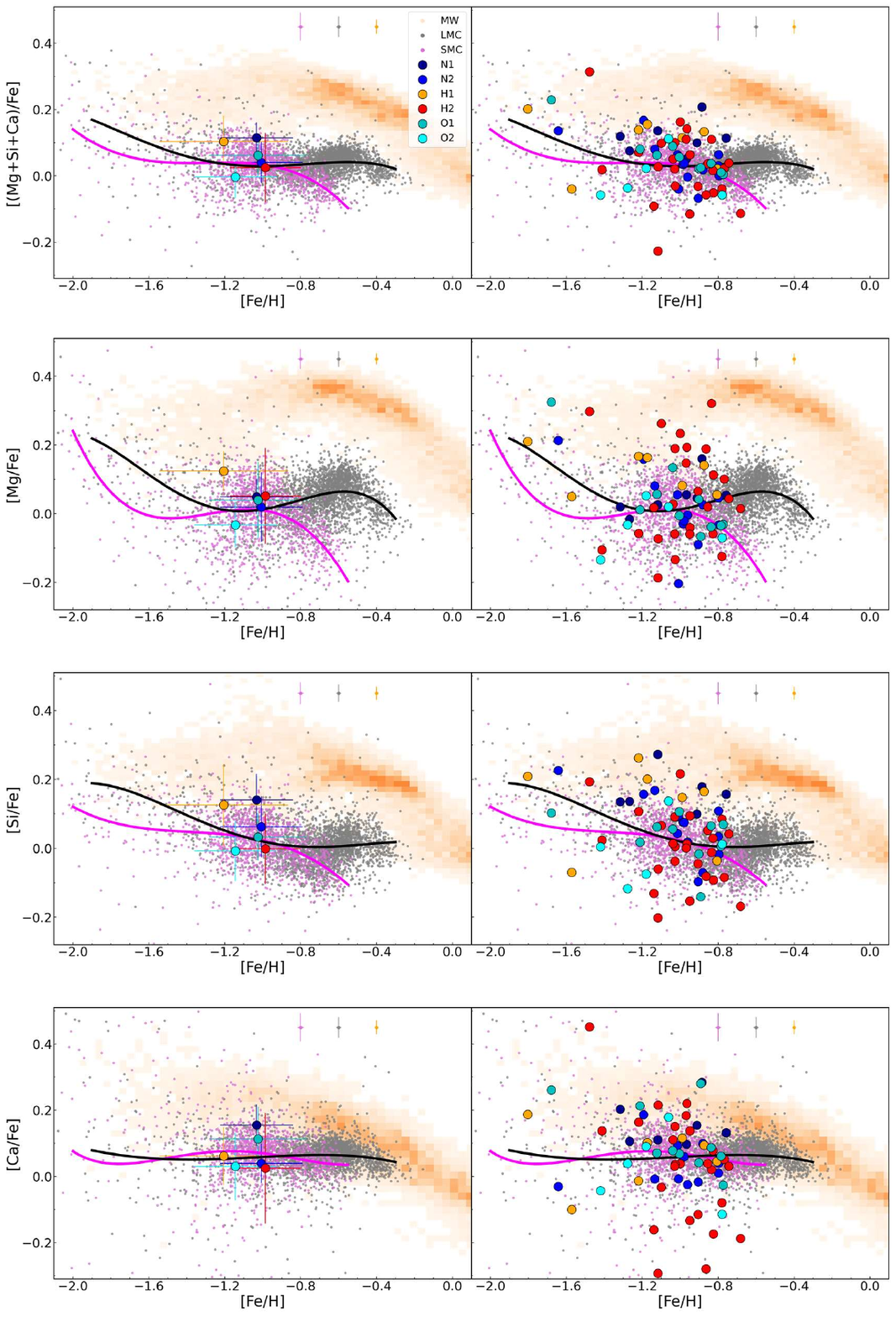}
\caption{The distribution of $\alpha$-abundances versus metallicity 
for the LMC (small grey dots) and SMC (small magenta dots). Also included are the density of  Milky Way stars  (orange)
and the six substructures analyzed in this study (filled 
colored circles with error bars which represent  the standard deviation of each region). Left panels show the means of our substructure fields while right panels show the individual stars in each field. In the top-right of each panel the representative errors for LMC, SMC and MW stars from APOGEE pipeline (ASPCAP) are also shown. The solid  black line and the solid magenta line represent the best polynomial fit for the LMC and SMC stars respectively.}
\label{alpha}
\end{figure*} 
 

\begin{figure*}[!ht]
\centering
\includegraphics[width=0.87\hsize,angle=0]{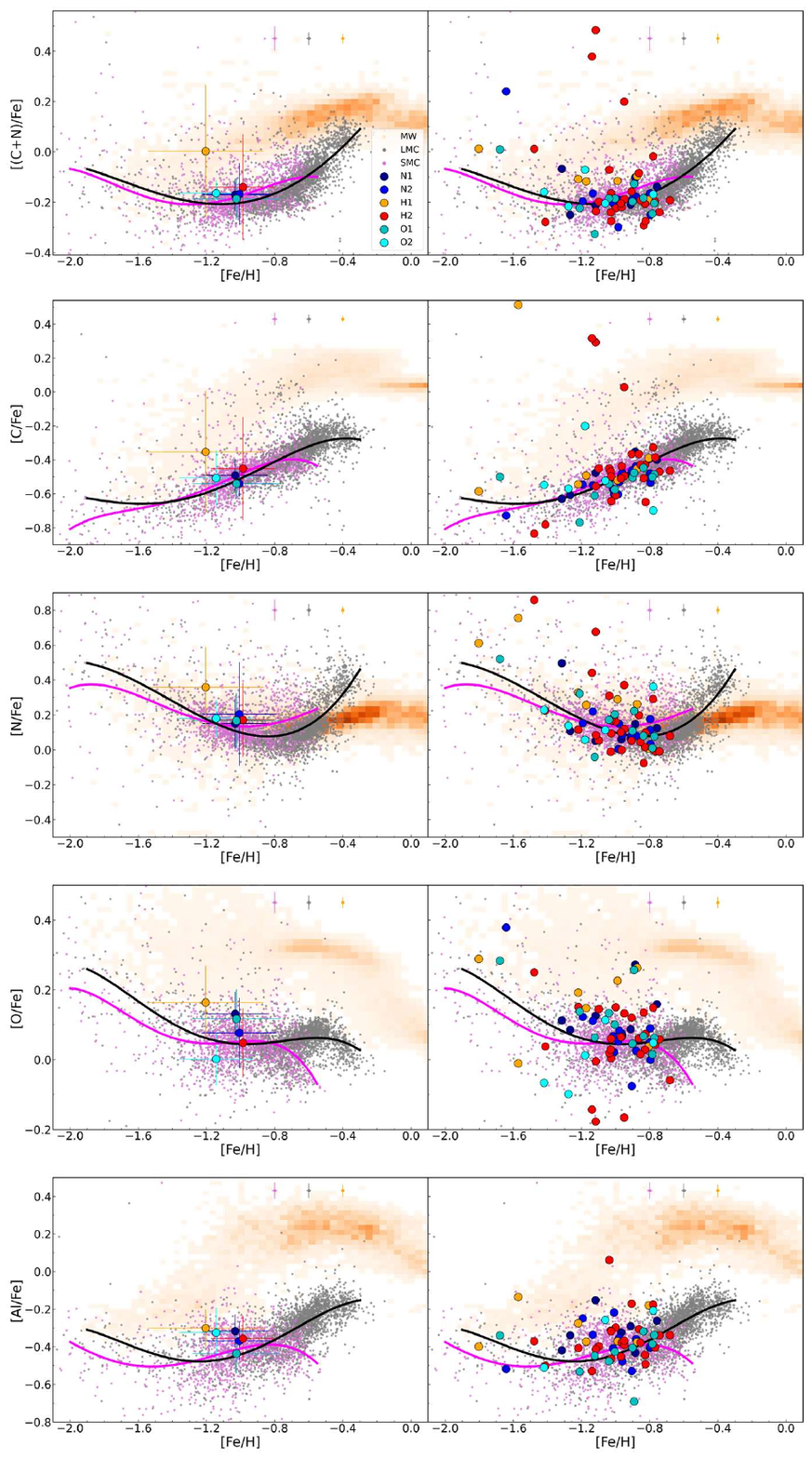}
\caption{
Similar to Figure \ref{alpha}, but for the distributions of
[C/Fe], [N/Fe], [O/Fe], [Al/Fe] versus [Fe/H]. 
}
\label{light}
\end{figure*} 
 
\begin{figure*}
\centering
\includegraphics[width=0.55\hsize,angle=0]{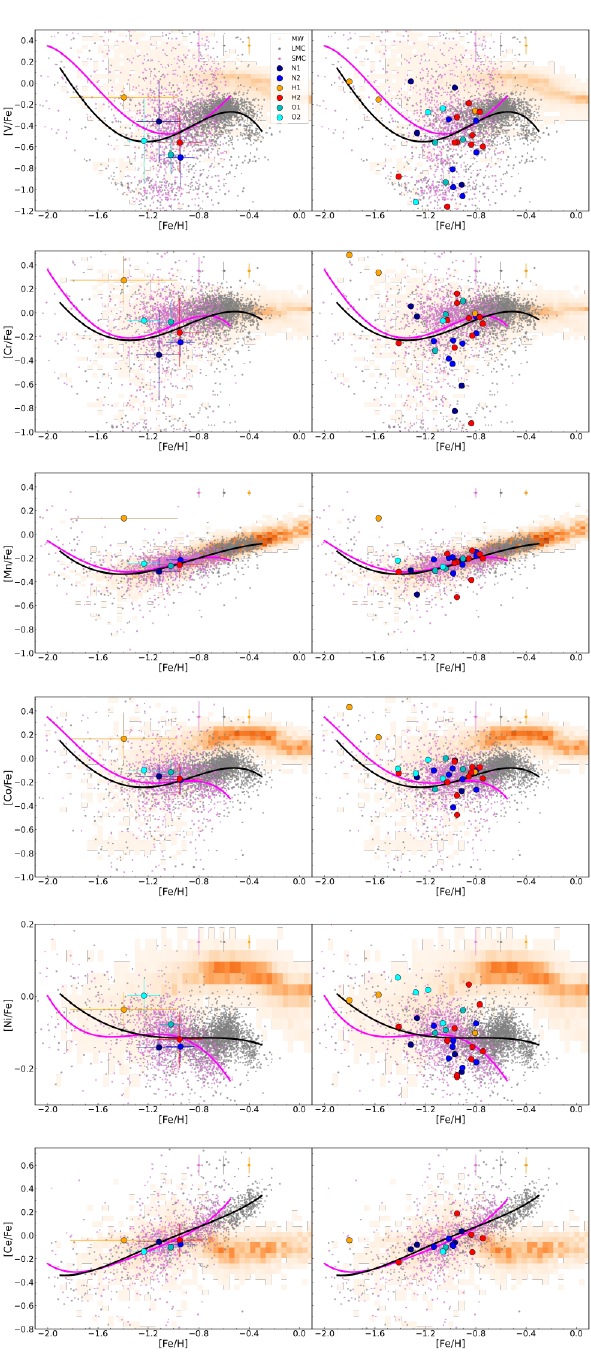}
\caption{
Similar to Figure \ref{alpha}, but for the distributions of [V/Fe], [Cr/Fe], [Mn/Fe], [Co/Fe], [Ni/Fe], [Ce/Fe] versus [Fe/H].
}
\label{iron}
\end{figure*}


\section{Results}
\label{sec:results}

\subsection{Metallicity distribution functions and radial profiles}
\label{subsec:mdfs}

In Figure \ref{MetDensity}, we show the metallicity probability distribution function using kernel density estimation (KDE) for the LMC, the SMC and the six periphery regions analyzed in this study. For the LMC, we subdivide the function into two  
parts: the inner region ($R_{\rm LMC}$ $<$ 6\degr) and 
the outer region (6\degr $\leq$ $R_{\rm LMC}$ $\leq$ 11\degr). 
There is a small offset in the peak of the outer LMC metallicity compared with that for the inner LMC.
This is expected due to the radial metallicity gradient in the LMC, behavior that has been studied previously by several authors
(\citealt{Cioni2009}; \citealt{Feast2010}; \citealt{Choudhury2021}; Povick et al.\ 2023b, in prep.).
The outer region is about 0.1 dex more metal poor than the inner region; this may be because the outer stars are older, as some photometric studies have found in this area of the LMC \citep{Piatti2013,Gatto2020,Nidever2019}.
The peak in metallicity of the SMC is about $-$1.1 dex. The observed 0.5 dex difference between the two galaxies is 
in agreement with that found by other studies that determine the metallicity of both MCs \citep[see, e.g., ][]{Noel2009, Meschin2014}.
Figure \ref{MetDensity} also shows that the foreground MW halo stars (Section \ref{subsec:mwforeground}
)
have a higher metallicity than stars in the Magellanic system, including those in the six substructure regions, with a peak at about $-$0.1 dex, but a wide range in metallicity values. 
Finally, we find that
five of our substructure regions show a peak in metallicity at about $-$1.0 dex, in agreement with the SMC's metallicity peak. However, region O2, with a peak at about $-$1.3, shows an Metallicity Distribution Function (MDF) that is significantly different;
we analyze this difference in more detail in the next section.



It is of   particular interest  to understand whether the stars in the six peripheral regions originated in the LMC or SMC.  One way to gain insight into this question is to see whether the mean chemical abundances of the stars in these fields are more consistent with following radial abundance trends from the LMC versus the SMC.
Figure \ref{prof_LMC_SMC} presents the radial metallicity and $\alpha$-abundance ([(Mg+Si+Ca)/Fe]) profiles as a function of the distance from the center of the LMC and SMC,
respectively, using the LMC and SMC central data points as well as those in the six outskirts regions. We highlight that this is the first time the metallicity and $\alpha$-abundance  profiles of the MCs are presented out to these distances, $\sim$20 deg from the LMC center. 

In this study we decided to use [(Mg+Si+Ca)/Fe] as alpha abundance instead of the [$\alpha$/M] value determined by ASPCAP. The reason for this is twofold: on one hand, we compare our  $\alpha$-abundance with the results found in \citet{Nidever2020}, who used ([(Mg+Si+Ca)/Fe]) as $\alpha$-abundance.  On the other hand, \citet{meszaros13} found a correlation between [$\alpha$/M] and [M/H] in ten Globular Clusters in APOGEE DR10 which still exists in DR16, according to 
 \citet{Nidever2020}. This correlation is mainly due to the fact that the second generation of stars is dominated by strong Al lines. \citet{Nidever2020} showed that the correlation is removed by using abundances relative to Fe instead of M.


We  find that
the LMC metallicity decreases with angular radius (see Figure \ref{prof_LMC_SMC}) from 0 out to $\sim$11 deg from the LMC center, with a slope of $-$0.03 dex deg$^{-1}$
taking into account the median value of each region from the main LMC body, which appears as filled grey symbols with error bars in Figure \ref{prof_LMC_SMC}.
This behavior is supported by several studies that have previously shown
that the older and more metal poor populations of
the LMC are mostly in its outskirts
\citep{Piatti2013,Carrera2011,Gatto2020}. Our results are also consistent with those reported by  \citet{Majewski2009}, who observed a similar metallicity gradient in the LMC but with a smaller sample size.
Regarding the six outer regions of interest here, in general, we observe that they match well 
the expected metallicity gradient of the LMC, 
especially regions N1, N2 and O1, despite the fact 
that these regions go beyond the farthest LMC main-body APOGEE stars (see the dashed line in Figure \ref{prof_LMC_SMC}, which shows the extrapolation of the metallicity gradient for LMC). Region H1 
most deviates from the LMC metallicity profile, yet its larger error bars also overlap  the extrapolation of the
LMC main body.

The $\alpha$-abundance profile, which we define here as given by the radial gradient of [(Mg+Si+Ca)/Fe], 
increases slightly for the LMC from 0.02 up to 0.1 dex values at larger radii, taking into account the median value of each region.  This is consistent with older (and $\alpha$-rich) populations dominating the outer LMC. Once again the regions N1, N2 and O1, exhibit a good match to this trend. On the other hand the discrepancy of region O2, and possibly H1 (which again has large error bars that include agreement with the extrapolation), is notable. Because the chemical evolution of the $\alpha$-element abundance is strongly influenced by the star formation history (SFH) of a galaxy, this discrepancy might suggest that O2 had a different chemical evolution from that of the LMC.

Figure \ref{prof_LMC_SMC} also shows the same radial metallicity and $\alpha$-abundance profiles for the SMC (right column). We observe that the SMC metallicity decreases sharply, with a slope of $-$0.04 dex deg$^{-1}$. In this case, we do not observe a good match between the extrapolated radial metallicity trend and the metallicities of stars in the regions in the outskirts, which are significantly more metal-rich than the extrapolated trend. However, we do not expect the metallicity trend to follow to those large SMC-centric
distances, since a galaxy like the SMC is not expected to have stars out to such large
radial distances (about 35\dgr or $\sim$44 kpc from the SMC center, see  right panel Figure \ref{prof_LMC_SMC}), considering that its
tidal radius is expected to be only $\sim$5.0 kpc \citep[$\sim4.5$ deg;][]{Massana2020}.  Note that \citet{Nidever2011} found an old intermediate-age population at a distance of about  10\dgr from the SMC center that are likely extratidal stars but could also be a bound stellar halo. \citet{Massana2020} also uncovered a tidally 
disrupted stellar feature that reaches as far out as 12\dgr from the SMC centre.
Were the
stars of interest here at very large radii 
stripped SMC stars, 
they would be expected to have
metallicities more like those of the SMC at $\sim$6\degr.
Only the median metallicities of the stars in H1 and O2 are close to being consistent with that hypothesis.
We also note that the inner negative metallicity gradient displayed by the SMC shows signs of stopping and even reversing sign in the outer regions, around 4\degr, near the limit of the APOGEE data \citep{Parisi2022}.

Regarding the $\alpha$-abundance profile, as in the case of the LMC, we 
note an increase of the $\alpha$ content with distance from the center of the SMC. In this case, 
there is
a good match for H2, O1 and H1 to the extrapolated trend. It is important to note that the observed radius of the SMC is about 11 kpc \citep{Nidever2011} from the center, therefore, the comparison with the most extreme regions like N1 and N2 does not make physical sense unless these regions were very strongly stripped.

\subsection{Chemical abundance patterns}

Figures \ref{alpha},  \ref{light}  and  \ref{iron} show 14 different elemental abundances as a function of metallicity for the stars in the substructures, colored as in the previous figures and as indicated in each figure, as well as for the main body of the LMC (grey dots) and SMC (magenta dots). Table \ref{tab:abundances} lists the mean and standard deviation of the 14 APOGEE chemical abundances analyzed for each of the six substructure fields. In each panel in Figures \ref{alpha}--\ref{iron}, the solid black and magenta lines show the best polynomial fit for the LMC and SMC, respectively. In what follows, we analyze the chemical abundances per region and compare them with the LMC, SMC and MW chemical abundances. We note that the determination of [V/Fe] abundances through the ASPCAP pipeline may lack precision and could be subject to biases in some cases, as evidenced by \citet{Hayes2023}. Therefore, analyses that include this element should be approached with caution.

We included  the [(C+N)/Fe] abundance in figure \ref{light} . This  abundances in the LMC tends to decrease   in the range -2.2<[Fe/H]<-1.2 , as observed and also reported  by \citet{Hasselquist2021}, and then starts to increase. A Similar pattern is observed  in the case of the  SMC. In the next sub-section, we analyze this abundance alongside light elements for each substructures.

\subsubsection{Regions N1 \& N2 }
\label{subsec:n1n2}

Region N1 is one of the (projected)  farthest of our six MC APOGEE regions
from the center of the LMC (see Figure \ref{prof_LMC_SMC}) and is located on the one of the arm-like features discovered by \citet{Mackey2016} in the north of the LMC (see Figure \ref{MapCoord}).  

The chemical abundance patterns of the APOGEE stars in this region are (out of the six) the most like those of the LMC. N1 exhibits the smallest metallicity spread ($\sigma$({\small{}{[}Fe/H{]} }=0.19 dex) of all six regions (see Table \ref{tab:param1}). Moreover, Figure \ref{prof_LMC_SMC} shows 
that the median [Fe/H] 
of the APOGEE stars in N1 ([Fe/H]=$-$1.03) follows the extrapolated trend of the LMC metallicity profile (represented by the straight line), taking into account the scatter of the data for this region. This finding in metallicity is in agreement with \citet{Majewski2009} who found a median  metallicity for the outer LMC population at radii of 15--20\dgr from the LMC center of [Fe/H]$\sim$ $-$1.0,  although with a large spread in metallicity. 
The median N1 metallicity of $-$1.03 dex matches the predicted value from the Majewski et al.\ metallicity gradient in the LMC periphery at the radius of N1. Moreover, the N1 metallicity 
matches the 
[Fe/H] $\approx$ $-$1.0 value derived from isochrone fitting of deep SMASH photometry in the LMC periphery by \citet{Nidever2019}. The association of N1 with the LMC is also suggested
by the probability density functions (see Figure \ref{MetDensity})
where N1 shows a shift in the metallicity distribution towards lower values than the outer LMC population shown there (from 6 to 11\degr), again in agreement with the LMC metallicity gradient. 



The median value of the [$\alpha$/Fe] for N1, represented as the largest color circles with error bars in Figure \ref{prof_LMC_SMC},
also shows a good match 
to the radial LMC extrapolated trendline (dashed black line) in Figure \ref{prof_LMC_SMC}. 


We find that the N1 $\alpha$-element abundances (Mg, Ca, and Si) as a function of metallicity (Figure \ref{alpha}) show good agreement with the LMC and SMC trends and are not consistent with the MW abundances trend. 
We measured the mean orthogonal distance of the stars in N1 to the best fit to the trendlines for the LMC (black line) and SMC (magenta line) and they show similar values (see Tables \ref{tab:Fit_LMC} and \ref{tab:Fit_SMC}).  
In our analysis of the orthogonal distance, we adopt a convention where distance above the curve are considered positive, while distance below are considered negative.This convention helps differentiate and analyze of these stars  relative to the curve, which represents the best fit.
In particular, for the $\alpha$ abundance [(Mg+Si+Ca)/Fe], the mean value for the orthogonal distance to the LMC trend is $-$0.097 dex and $-$0.089 dex for the SMC. In Figure \ref{alpha}, we only observe a slight deviation of the median value for [Si/Fe] in N1, in comparison with the other regions in our analysis.



This conclusion regarding the abundance patterns of N1 also generally holds for other light as well as iron-peak elements, for which we show the
LMC and SMC trends for our data in Figures \ref{light} and \ref{iron}, respectively.   
More specifically, the trends for light elements as a function of metallicity show the stars in N1 to have a good match with the overall trends for the LMC, for the case  (C+N), C and  N (Figure \ref{light}). For Al and O we observe  offsets, with similar values (See Tables \ref{tab:Fit_LMC} and \ref{tab:Fit_SMC}), from the LMC and SMC fits. However, the separation from the MW trend is clear, indicating a better match with the MCs.
The case of C, O and Al 
are especially useful, because for these abundance ratios there is 
a clear difference between the trends of the LMC and SMC and that of 
the MW. In contrast, the trend of [N/Fe] as a function of metallicity shows an overlap between the LMC, SMC and MW within the metallicity range of interest in this work (i.e., 
[Fe/H] $<$ $-0.6$). For which includes all the six outskirt regions analyzed in this work. For C, O and Al the abundance ratio for N1 is lower by $\sim$0.35 dex with respect to the MW, and is much closer to the trend for the LMC and SMC. 

Finally, for the Fe-peak elements (Figure \ref{iron}), we again observe a similarity of the N1 abundance relations to those found in the LMC and SMC trends. Although, based on the mean of the orthogonal distances of the stars  to the best fit for LMC and SMC, the agreement appears to be slightly better with the LMC (see Tables \ref{tab:Fit_LMC} and \ref{tab:Fit_SMC}).

In the particular case of V, Co and Ni, there is a clear difference in their trend for LMC and SMC with respect to the MW trend, and so those are particularly useful elements for discriminating MC substructures from MW contaminants.
This is in contrast to the abundance ratios involving Mn and Cr, where there is 
an overlap in the abundance values as a function of metallicity among all the galaxies shown. 
In any case, the abundance ratios for N1 are found to be consistent with the LMC and SMC trends for all of these iron-peak elements, taking into account the scatter of the data, especially for the case of Ni and Co, where there is a significant difference between MCs trends in comparison with the MW.





Of course, because of the general similarity of the SMC and LMC abundances over the metallicities of interest, any similarity of the N1 field to the LMC trendlines implicitly suggests a similarity also to the SMC trendlines.
However, the position and distance of the N1 field 
from the SMC (see Figures \ref{MapCoord} and \ref{prof_LMC_SMC}) as well as the radial velocity (RV) observed for the APOGEE stars in this region (see next section) together are simply incompatible with those of the SMC. 


The analysis for N2 is almost identical to N1, as expected due to the proximity of these two fields. Both are part of the same northern LMC arm and share similar proper motions (see Table \ref{tab:param1} and Figure \ref{MapCoord}).
Furthermore, N2 has a mean metallicity similar to that of N1, $\langle$[Fe/H]$\rangle$ = $-$1.01 (see Table \ref{tab:param1} and Figure \ref{MetDensity}), and shows a good agreement with the LMC in the radial metallicity and $\alpha$ gradients presented in Figure \ref{prof_LMC_SMC}. In our analysis presented in Figures \ref{alpha},  \ref{light} and \ref{iron}, 
N2 shows excellent agreement with the LMC for the light, $\alpha$, and Fe-peak elements.  Finally, the analysis of cerium (Ce) for N1 and N2 reveals a strong concordance with the LMC and SMC, characterized by minimal dispersion around the trend-lines for the MCs.

\subsubsection{Regions H1 \& H2}
\label{subsec:h1h2}
H1 is one the regions closest to the center of the LMC, at a galactocentric distance of about $\sim$11.5\dgr (see Figure \ref{prof_LMC_SMC}), and lies in 
the southeast LMC periphery (see Figure \ref{MapCoord}). The proper motions of its MC-related stars show good agreement with the LMC.
and C22 concluded that their kinematics are associated with the LMC outer disk.
Also, 
this region is the most metal poor ([Fe/H]=$-$1.25) and shows the largest spread in metallicity among the six regions ($\sigma$({\small{}{[}Fe/H{]} }=0.34 dex), as listed in Table \ref{tab:param1}. This is also reflected in the LMC radial metallicity profile (see Figure \ref{prof_LMC_SMC}), where H1 has a median metallicity most separated from the trend represented by the straight line.
However, the large error bars fall within the extrapolation of the LMC gradient at H1's distance.
For the radial $\alpha$-abundance profile in Figure \ref{prof_LMC_SMC}, we find that H1 median value is slightly above the LMC trend represented by the straight light, but, again, as in the metallicity profile, it is in agreement within the error bars. We note that the H1 median metallicity and $\alpha$-abundance would not match the extrapolated values for SMC trends, considering that stars at those distances (about $\sim$20\dgr from the SMC's center) should be extratidal stars, if they were of SMC origins and their abundances should match those at about $\sim$6\dgr from the SMC's center.

In the case of the $\alpha$-elements O, Mg, Si and [(Mg+Si+Ca)/Fe], we find the H1 mean value to be higher than that of the LMC abundance distribution trend, but the trend is just within the error bar for H1 (Figures \ref{alpha} and \ref{light}). We note that the individual stars have a wide [Fe/H] range in H1, and we can observe that one or two stars (depending on the elements) out of the 7 stars in H1 are right on or below the LMC trend for these elements.
Ca is the $\alpha$-element that shows the best agreement with the LMC (see Figures \ref{trend_comp_Nid} and \ref{alpha}). 


With respect to the light elements  (C+N), C and N (see Figure \ref{light}), the median values for the H1 stars
are higher than that for the LMC trend, 
and a similar behavior is seen for the Fe-peak elements where for the H1 stars the V, Cr, Mn and Co abundance ratios are higher
not only in comparison to the other regions, but even when contrasted with the MW (see Figure \ref{iron}). Regarding cerium (Ce), we observe a slightly elevated value for H1 (only one star) in comparison to the trend-lines of the MCs.
We note, however, that in some cases the number of stars with measurements of these elements and with S/N $>$ 60 is  only between one and three stars.




H2 is the closest region to both the centers of the LMC and the SMC, among the six regions of interest, at $\sim$11\dgr from each galaxy (see Figure \ref{prof_LMC_SMC}) and it has the highest number of member stars of the six regions (27 stars).  The Vector Point diagram of the proper motions
(Figure \ref{Proper_Motion}) shows that most of the H2 stars are in between the distributions of LMC and SMC stars. Thus, its member stars, due to their distance to both the SMC and LMC center  (at about 11\dgr from each) and  their PM
could be related to either galaxy. 
However,
\citet{Nidever2011} determined a projected galactocentric distance of 10 kpc (10\degr) as the limit
for SMC membership. Therefore, if belonging to the SMC, H2 could be an extreme, outlying SMC field or it could contain extratidal stars from the perturbed SMC. 


 
The mean metallicity of H2 is [Fe/H] = $-$1.0, which, at the position of H2 on the sky, is consistent with the radial metallicity gradients of  the LMC and  a slightly  with the  SMC, taking into account  the dispersion of H2.
(see Figure \ref{prof_LMC_SMC}). However, the probability density function for H2 in Figure \ref{MetDensity} shows a slightly better fit with the SMC profile than with either LMC profile.
In the case of the light elements (Figure \ref{light}), we observe a good match with the LMC and SMC trends for most of the H2 stars. However, a couple of stars show a significant discrepancy, especially for (C+N), C and N, which  show about 0.6 dex difference from the LMC and SMC trends (see Figure \ref{light}).
For the $\alpha$-elements (Figure \ref{alpha}) there are significant spreads for [Mg/Fe] and [Ca/Fe], the largest spreads among all six of the regions (Table \ref{tab:abundances}).

 
On the other hand, the H2 abundance ratios for 
Fe-peak elements (Figure \ref{iron})  show good agreement
with those for both the LMC and SMC (of course, for the stars having measurements), especially for the elements where there is a clear difference between the MW and MCs, that is for
V, Co and Ni. For Mn and Cr, as we noted
previously (Section \ref{subsec:n1n2}), there is a strong
overlap among the MW, LMC and SMC distributions. For Ni do we observe a couple of stars with extreme values, low and high in comparison with the LMC trend.
Also, We observe  two stars in H2, specifically for  Cr and V, that exhibit very low abundances. However, these abundances are still in agreement with those found in the LMC and SMC (See figure \ref{iron}). The analysis of cerium (Ce) for H2 also shows a good compatibility with the MCs, although we only have five stars in this field for this element.

\begin{table*}

\caption{Mean and standard deviation of fourteen APOGEE chemical abundances for the six substructure fields }
\label{tab:abundances} 
\begin{center}
\begin{tabular}{lllllllllllll}
\hline 
\multicolumn{1}{l|}{\textbf{}} & \multicolumn{12}{c}{\textbf{Regions}}                                                                                                           
                         \\ \cline{2-13} 
\multicolumn{1}{l|}{Element}         & \multicolumn{2}{c|}{N1}    & \multicolumn{2}{c|}{N2}    & \multicolumn{2}{c|}{H1}    & \multicolumn{2}{c|}{H2}    & \multicolumn{2}{c|}{O1
}    & \multicolumn{2}{c}{O2} \\
\multicolumn{1}{l|}{}            &  Mean & \multicolumn{1}{l|}{$\sigma$ } & Mean & \multicolumn{1}{l|}{$\sigma$} & Mean & \multicolumn{1}{l|}{$\sigma$} & Mean& \multicolumn{1}{l|}{$\sigma$} & Mean & \multicolumn{1}{l|}{$\sigma$} & Mean  & $\sigma$  \\ 
\hline
$[$(C+N)/Fe$]$  &   $-$0.17  & 0.06 & $-$0.17 &  0.12  & 0.00  &  0.26    & $-$0.14  &  0.21  & $-$0.19  & 0.08 &   $-$0.16  &  0.05  \tabularnewline
$[$C/Fe$]$  &   $-$0.49  & 0.09 & $-$0.54 &  0.07 & $-$0.35  &  0.36    & $-$0.45  &  0.31  & $-$0.54   & 0.09 &   $-$0.46  &  0.15  \tabularnewline
$[$N/Fe$]$   &   0.15     & 0.14 & 0.20   &  0.30 &  0.36  &  0.23    & 0.22  &   0.30  & 0.17   & 0.16&   0.13    &  0.06      \tabularnewline
$[$O/Fe$]$   &   0.13     & 0.06 & 0.08   &  0.10 &  0.16  &  0.10    & 0.06  &   0.11  & 0.12  & 0.08&   $-$0.01    &  0.08     \tabularnewline
$[$Al/Fe$]$ &   $-$0.32  & 0.09 & $-$0.37 &  0.10 &  $-$0.30  &  0.10    & $-$0.37  & 0.12     & $-$0.44   & 0.11 &   $-$0.38(2) &  0.13  \\
$[$Mg/Fe$]$ &   0.05     & 0.05 & 0.02  &  0.10 &  0.12  &  0.06    & 0.13  &   0.17 & 0.04   & 0.11 &   $-$0.02    &  0.07   \\
$[$Si/Fe$]$  &  0.14     & 0.08 & 0.06  &  0.09 &  0.13  &  0.12    & 0.07  &   0.07  & 0.03   & 0.07 &   $-$0.01    &  0.10   \\
$[$Ca/Fe$]$  &  0.16     & 0.06 & 0.04  &  0.06 &  0.06  &  0.09    & 0.03 &   0.17 & 0.11   & 0.10 &  0.07  &  0.08    \\
$[\alpha$/Fe$]$\tablefootmark{1} &  0.12    & 0.05 & 0.04  &  0.06 &  0.10  &  0.08    & 0.02  &   0.10  & 0.06  & 0.06 &  0.00  &  0.10  \tabularnewline
$[$V/Fe$]$   &  $-$0.36(4) & 0.39 & 0.70(6) &  0.28 &  0.14(3)  &  0.11    & $-$0.56(10)  &  0.27  & $-$0.67(3)  & 0.18 &   $-$0.54(3)&  0.41    \\
$[$Cr/Fe$]$  &  $-$0.35(4) & 0.37 & 0.25(7) &  0.12 &  0.27(3)  &  0.21    & $-$0.16(10)  &  0.29  & $-$0.08(3) & 0.18 &   $-$0.07(1)    &  0.00    \\
$[$Mn/Fe$]$  &  $-$0.31(4) & 0.12 & $-$0.22(7)  &  0.06 &  0.14(1)  &  0.00   & $-$0.26(10)  &  0.11 & $-$0.27(3)   & 0.04 &   $-$0.25(2)    &  0.03    \\
$[$Fe/H$]$   &  $-$1.03     & 0.19 & $-$1.01   &  0.22 &  $-$1.25  &  0.34    & $-$1.00  &   0.22  & $-$1.03  & 0.26 &  $-$1.14    &  0.22    \\
$[$Co/Fe$]$  &  $-$0.15(3)  & 0.11 & $-$0.18(7)  &  0.11 &  0.16(3)  &  0.22    & $-$0.17(10) &  0.13  & $-$0.12(3)  & 0.11&   $-$0.10(4)    &  0.06    \\
$[$Ni/Fe$]$  &  $-$0.14(4)  & 0.05 & $-$0.14(7)  &  0.04 &  $-$0.04(3) &  0.05    & $-$0.12(10)  &  0.08  & $-$0.08(3)  & 0.03 &   0.00(4)    &  0.05    \\
$[$Ce/Fe$]$  &  $-$0.06(4)  & 0.06 & $-$0.08(4)  &  0.03 &  $-$0.04(1) &  0.00    & $-$0.04(5)  &  0.14  & $-$0.10(1)  & 0.00 &  $-$0.14(1)   &  0.00   \\
\hline
\end{tabular}\\
\tablefoot{
\tablefoottext{1}{\small{}{[}$\alpha$/Fe{]}:[(Mg+Si+Ca)/Fe] }
}
\end{center}

\end{table*}
\begin{table*}
 \begin{center}
\caption{The mean of the minimum orthogonal distance  for stars in  each of the substructures to the best-fit of the LMC in solid black line in Figures \ref{light},\ref{alpha} and \ref{iron}.}
\label{tab:Fit_LMC} 
\begin{tabular}{ccccccc}
\hline 
\hline
Element &N1& N2& H1& H2 &O1 &O2 \\ 
\hline

$[$C/Fe$]$&	0.023&	$-$0.037 &	0.206 &	0.046 &	$-$0.035 &	0.051 \\
$[$N/Fe$]$&	0.018&	0.079 &	0.153 &	0.079 &	0.031 &	0.010 \\
$[$O/Fe$]$&	0.075&	0.017 &	0.082 &	$-$0.001 &	0.052 &	$-$0.064 \\
$[$Mg/Fe$]$&	0.025 &	$-$0.011 &	0.082 &	0.025 &	0.004 &	$-$0.058 \\
$[$Al/Fe$]$&	0.109 &	0.047 &	0.103 &	0.047 &	$-$0.023 &	0.092 \\
$[$Si/Fe$]$&	0.108 &	0.033 &	0.073 &	$-$0.030 &	0.000	& $-$0.056 \\
$[$Ca/Fe$]$&	0.097 &	$-$0.020 &	0.009 &	$-$0.030 &	0.054 &	$-$0.025 \\
$[$V/Fe$]$&	0.198 &	$-$0.075 &	0.092 &	0.091 &	0.194 &	0.015 \\
$[$Cr/Fe$]$&	$-$0.120 &	$-$0.089 & $-$0.035 &	$-$0.046 &	$-$0.058 &	0.098 \\
$[$Mn/Fe$]$&	$-$0.070 &	0.024 &	$-$0.086 &	$-$0.045 &	$-$0.052 &	0.024 \\
$[$Co/Fe$]$&	$-$0.003 &	0.013 &	0.170 &	$-$0.093 &	$-$0.047 &	0.130 \\
$[$Ni/Fe$]$&	$-$0.056 &	0.008 &	$-$0.008 &	0.013 &	$-$0.008 &	0.069 \\
$[$Ce/Fe$]$&	0.260 &	0.385 &	0.328 &	0.258 &	0.415 &	$-$0.028 \\
$[$(Mg+Si+Ca)/Fe$]$&	$-$0.097 &	$-$0.103 &	0.035 &	0.005 &	$-$0.085 &	$-$0.031 \\

\hline 
\end{tabular}
\end{center}
\end{table*}
\begin{table*}
\caption{The mean of the minimum orthogonal distance  for each substructures  to the best-fit of the SMC  in solid magenta line in Figure \ref{light}, \ref{alpha} and \ref{iron}.}
\label{tab:Fit_SMC} 
\begin{center}
\begin{tabular}{ccccccc}
\hline 
\hline
Element &N1& N2& H1& H2 &O1 &O2   \tabularnewline
\hline
$[$C/Fe$]$&	0.009&	$-$0.0051	&0.195&	0.037&	$-$0.045&	0.034\\
$[$N/Fe$]$&	$-$0.008&	0.047&	0.158&	0.051&	0.002&	0.005
\\
$[$O/Fe$]$&0.082&	0.021&	0.100&	0.006&	0.060&	$-$0.0048
\\
$[$Mg/Fe$]$&0.054&	0.0022&	0.128&	0.059&	0.045	&$-$0.026
\\
$[$Al/Fe$]$&	0.107&	0.044&	0.146&	0.052	&$-$0.018&	0.112
\\
$[$Si/Fe$]$&	0.120&	0.046&	0.100&	$-$0.015&	0.017&	$-$0.036
\\
$[$Ca/Fe$]$&	0.090&	$-$0.023&	0.006&	$-$0.034&	0.052&	$-$0.037
\\
$[$V/Fe$]$&	0.121&	$-$0.123&	$-$0.030&	0.043&	0.142&	$-$0.043
\\
$[$Cr/Fe$]$&$-$0.100&	$-$0.0119	&$-$0.051&	$-$0.060&	$-$0.089&	0.061
\\
$[$Mn/Fe$]$&$-$0.077	&0.016&	$-$0.100&	$-$0.052&	$-$0.061&	0.015
\\
$[$Co/Fe$]$&	0.001&	0.019&	0.1465&	$-$0.051&	$-$0.035&	0.091
\\
$[$Ni/Fe$]$&$-$0.048&	0.016&	0.004	&0.021&	0.003&	0.077
\\
$[$Ce/Fe$]$&	0.255&	0.359&	0.332&	0.231&	0.366&	$-$0.011
\\
$[$(Mg+Si+Ca)/Fe$]$&	$-$0.089&	$-$0.093&	0.048&	0.024&	$-$0.066&	$-$0.021
\\

\hline 
\end{tabular}
\end{center}
\end{table*}

\subsubsection{Regions O1 \& O2 }
\label{subsec:o1o2}

Of the six fields, O1 and O2 are sampling substructures located in the farthest southern portion of the LMC.  We have 10 selected stars in O1 and 5 
in O2.  The metallicity distribution functions of both regions (see Figure \ref{MetDensity}), also analyzed in C22\footnote{We note that there is a slight difference between our MDF and that of C22, due to the smaller sample presented in this paper. This is because our selection includes an extra criterion related to 
the S/N, which is required to analyze the chemical abundances.}, shows that O2 is slightly more metal poor than the SMC mean, but significantly more metal poor than the LMC mean, with a difference of about 0.6 dex in their mean values. 
This suggests a better match with the SMC than with the LMC.  A similar behavior is observed in O1, with a MDF in between the outer LMC and SMC values.  


Our analysis of
the radial metallicity profile for the LMC 
(Figure \ref{prof_LMC_SMC}) shows a relatively good match in the case of O1, but
a difference with the LMC gradient
for the case of O2.  A similar behavior is observed in the radial $\alpha$-abundance profile (Figure \ref{prof_LMC_SMC}).  O2 stars show a large difference in $\alpha$-abundance in comparison with the LMC trend, where extrapolation to O1 and O2 distances is shown as a dashed line in Figure \ref{prof_LMC_SMC}. When we place the results of these fields in the radial metallicity profile (see Figure \ref{prof_LMC_SMC}) from the SMC center, we find that region O2 shows the best match with the SMC metallicity radial trend among the six regions. In contrast, the median of region O1 shows a large difference.  However, the extrapolated values of SMC $\alpha$-abundance from its radial trend at the positions of O1 and O2 show a better match for O1 and a larger difference for O2.  

We can investigate further the origin of these fields by looking at their chemical abundance patterns.  It is interesting to note that region O1, in general, shows a good match in chemical patterns in comparison with both the LMC and SMC.  In the particular case of the $\alpha$-elements (see Figure \ref{alpha}), we observe good agreement as well, but especially with the SMC (see also the next Section \ref{sec:discussion} and Figures \ref{trend_comp} and \ref{trend_comp_Nid} explained there).  

The case of O2 is more difficult to analyze, mainly due to the small number of stars, only five.  Nevertheless, we observe an increase in the $\alpha$-elements starting from the most metal-poor star in O2, at [Fe/H=]$-$1.4 dex, the most clear example is for Mg and Ca. For the Fe-peak elements, we observe good agreements with the LMC and SMC taking into account the few members for this substructure.  We just note for Ni the most significant difference, but again in agreements with LMC and SMC. This behavior for Ni is similar to region H1, with similar mean value and scatter (see Table \ref{tab:abundances}).

Finally, the analysis of cerium (Ce) for O1 and O2 reveals a good agreement with LMC and SMC. However, it is important to note that there is only one star for this element in each region O1 and O2.



\begin{table*}

\centering

\caption{The mean and standard deviation of the minimum orthogonal distance for each substructure and each object (MW, LMC, SMC, Fnx, Sgr, Scl and GSE) represented in Figures \ref{orthogonal_distance_obj} for Mg and Si according to the trendline performed by \citet[see Figure \ref{trend_comp}]{Hasselquist2021}.}
\label{tab:dist_has_alpha} 
\begin{tabular}{lllllllllll}
\hline 
\multicolumn{1}{l|}{\textbf{}} & \multicolumn{10}{c}{\textbf{Objects}}                                                                                                           
                         \\ \cline{2-11} 
\multicolumn{1}{l|}{Regions}         & \multicolumn{2}{c|}{LMC}    & \multicolumn{2}{c|}{SMC}    & \multicolumn{2}{c|}{GSE}    & \multicolumn{2}{c|}{Sgr}    & \multicolumn{2}{c|}{Fnx
}   \\
\multicolumn{1}{l|}{}            & Mean & \multicolumn{1}{l|}{$\sigma$ } & Mean& \multicolumn{1}{l|}{$\sigma$} &Mean & \multicolumn{1}{l|}{$\sigma$} & Mean& \multicolumn{1}{l|}{$\sigma$} & Mean & \multicolumn{1}{l|}{$\sigma$}    \\ 

\hline
&&&&[Si/Fe]&&&&&\\
N1	&0.129&	0.075	&0.132&	0.076&	0.013&	0.082&	0.147&	0.083&	0.256&	0.076 \\ 
N2&	0.053&	0.071	&0.056&	0.075&	$-$0.054&	0.060&	0.080&	0.059&	0.178&	0.075 \\ 
H1&	0.087&	0.013	&0.100&	0.115&	$-$0.033&	0.127&	0.091&	0.136&	0.219&	0.117 \\ 
H2&	$-$0.009&	0.096	&$-$0.004&	0.095&	$-$0.116&	0.096&	0.018	&0.098&	0.118&	0.095 \\ 
O1&	0.020&	0.065&	0.026&	0.064&	$-$0.086&	0.070	&0.047&	0.074&	0.147&	0.064 \\ 
O2&	$-$0.031&	0.098&	$-$0.025&	0.096&	$-$0.158&	0.115&	$-$0.026&	0.117&	0.102&	0.093 \\ 
 \hline
&&&&[Mg/Fe]&&&&&\\
N1&	0.039&	0.066	&0.041&	0.064&	$-$0.076&	0.089&	0.059&	0.091&	0.166&	0.061\\
N2&	0.009&	0.084&	0.013&	0.092&	$-$0.096&	0.081&	0.038&	0.078&	0.135&	0.091\\
H1&	0.085&	0.061&	0.099&	0.050&	$-$0.034&	0.069&	0.090&	0.081&	0.217&	0.051\\
H2&	0.045&	0.140&	0.049&	0.141&	$-$0.061&	0.144&	0.074&	0.144&	0.172&	0.140\\
O1&	0.026&	0.078&	0.033&	0.087&	$-$0.079&	0.067&	0.053&	0.060&	0.153&	0.083\\
O2&	$-$0.056&	0.079&	$-$0.049&	0.070&	$-$0.182&	0.091&	$-$0.051&	0.094&	0.077&	0.068\\

\hline
\end{tabular}\\
\end{table*}
\begin{table*}
\centering
\caption{The mean and standard deviation of the minimum orthogonal distance for each substructures and each object (MW, LMC, SMC, Fnx, Sgr, Scl) represented in Figures \ref{orthogonal_distance_obj} and \ref{trend_comp_Nid} for $\alpha$-elements according to the trendline performed by \citet{Nidever2020}.}
\label{tab:dist_nid_alpha} 
\begin{tabular}{lllllllllllll}
\hline 
\multicolumn{1}{l|}{\textbf{}} & \multicolumn{12}{c}{\textbf{Objects}}                                                                                                           
                         \\ \cline{2-13} 
\multicolumn{1}{l|}{Regions}         & \multicolumn{2}{c|}{MW}    & \multicolumn{2}{c|}{LMC}    & \multicolumn{2}{c|}{SMC}    & \multicolumn{2}{c|}{Fnx}    & \multicolumn{2}{c|}{Sgr
}    & \multicolumn{2}{c}{Scl} \\
\multicolumn{1}{l|}{}            & Mean & \multicolumn{1}{l|}{$\sigma$ } & Mean& \multicolumn{1}{l|}{$\sigma$} &Mean & \multicolumn{1}{l|}{$\sigma$} & Mean& \multicolumn{1}{l|}{$\sigma$} & Mean & \multicolumn{1}{l|}{$\sigma$} & Mean         & $\sigma$       \\ 
\hline
N1&	$-$0.132&	0.050&	0.064&	0.045&	0.107&	0.046&	0.294&	0.064&	0.096&	0.071&	0.222&	0.114\\
N2&	$-$0.205&	0.055&	$-$0.014	&0.06&	0.026&	0.061&	0.234&	0.042&	0.026&	0.055&	0.119&	0.182\\
H1&	$-$0.152&	0.076&	0.036&	0.078&	0.078&	0.070&	0.209&	0.150&	0.035&	0.113&	0.128&	0.173\\
H2&	$-$0.219&	0.101&	$-$0.027&	0.103&	0.017&	0.102&	0.213&	0.096&	0.019&	0.095&	0.170&	0.107\\
O1&	$-$0.183&	0.048&	0.005&	0.049&	0.047&	0.048&	0.228&	0.050&	0.043&	0.017&	0.183&	0.079\\
O2&	$-$0.257&	0.066&	$-$0.054&	0.070&	$-$0.007&	0.060&	0.156&	0.100	&$-$0.047&	0.089&	0.064&	0.131\\

\hline
\end{tabular}

\end{table*}

\subsection{Kinematical analysis}

\begin{figure}
\centering
\includegraphics[width=3.2in,height=3in]{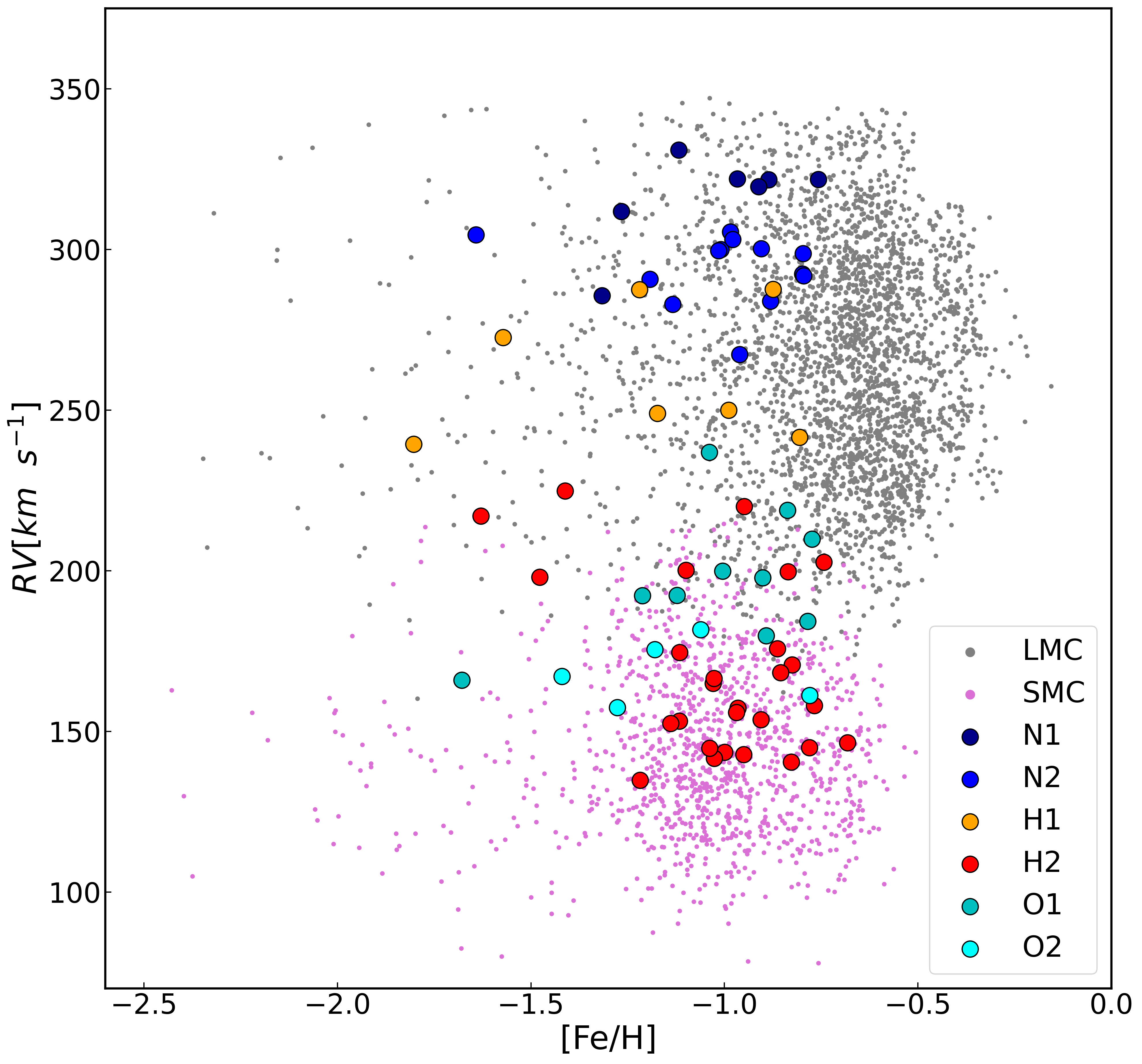}
\caption{Radial velocity versus metallicity of the stars in the six APOGEE MC periphery regions compared with LMC and SMC stars, with all data from APOGEE-2 DR17.}
\label{Met_Vel}
\end{figure} 

Figure \ref{Met_Vel} displays the heliocentric radial velocities (RVs) as a function of metallicity for all the stars in the substructures, as well as for LMC and SMC stars. From this figure, N1 and N2 share similar radial velocities ($\langle{\small{}RV_{H}}\rangle \sim$ 300 \kmse, see also Table \ref{tab:param1}), both of which are incompatible with those of the SMC. This is also the case for the stars in H1. In contrast, the bulk of H2 members have metallicities and RVs 
in agreement with those of 
SMC stars. The median radial velocity of H2 ($\langle {RV_{\rm H}}\rangle$=168.6 km s$^{-1}$) is the lowest median value among our fields (see Table \ref{tab:param1}) although with large scatter of $\sigma$({\small{}{[}RV$_{\rm H}${]} }) = 26.5 \kmse. 
However, as stated in the previous section, H2 is beyond the limit for SMC membership, which is determined at 10\dgr \citep{Nidever2011}. Therefore, if belonging to the SMC, H2 at could be an extreme, outlying SMC field or it could contain extratidal stars from the perturbed SMC. O1 and O2 members present RVs and metallicities close to the SMC values, but in between the two MCs, as we can be seen in Figure \ref{Met_Vel}. 

\begin{figure}
\centering

\includegraphics[width=3.2in,height=7in]{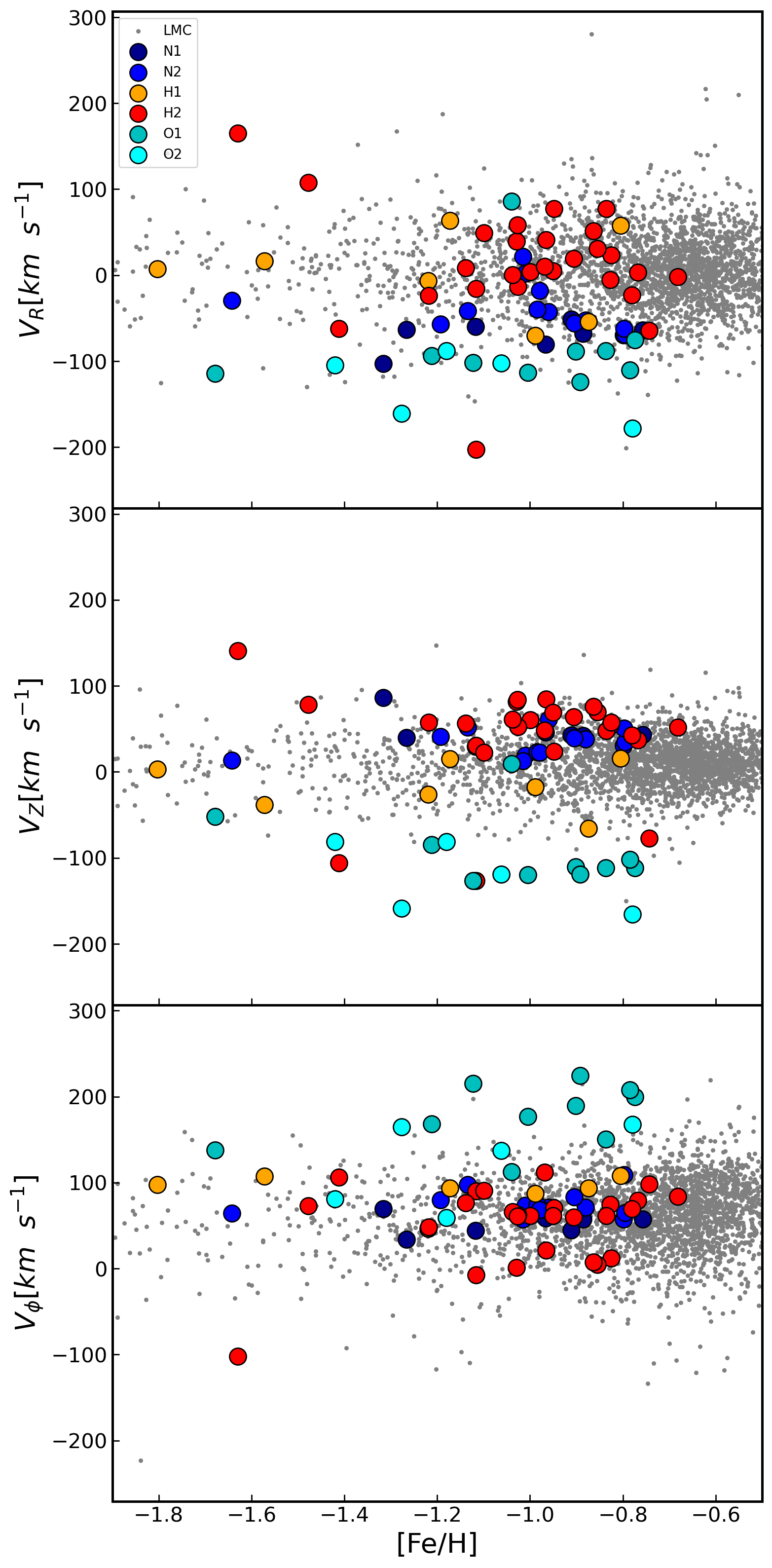}
\caption{Vertical velocity ($V_{z,{\rm LMC}}$), radial velocity ($V_{r,{\rm LMC}}$), and azimuthal velocity ($V_{\phi,{\rm LMC}}$) versus metallicity of stars in the six targeted substructure regions, and including stars targeted by APOGEE in the central LMC. Velocities were calculated by \citet{Cheng2022} with respect to the assumed LMC disk plane, as described in Section \ref{sec:data}.}
\label{vel_chen}
\end{figure}

\begin{figure}
\centering

\includegraphics[width=3.2in,height=7in]{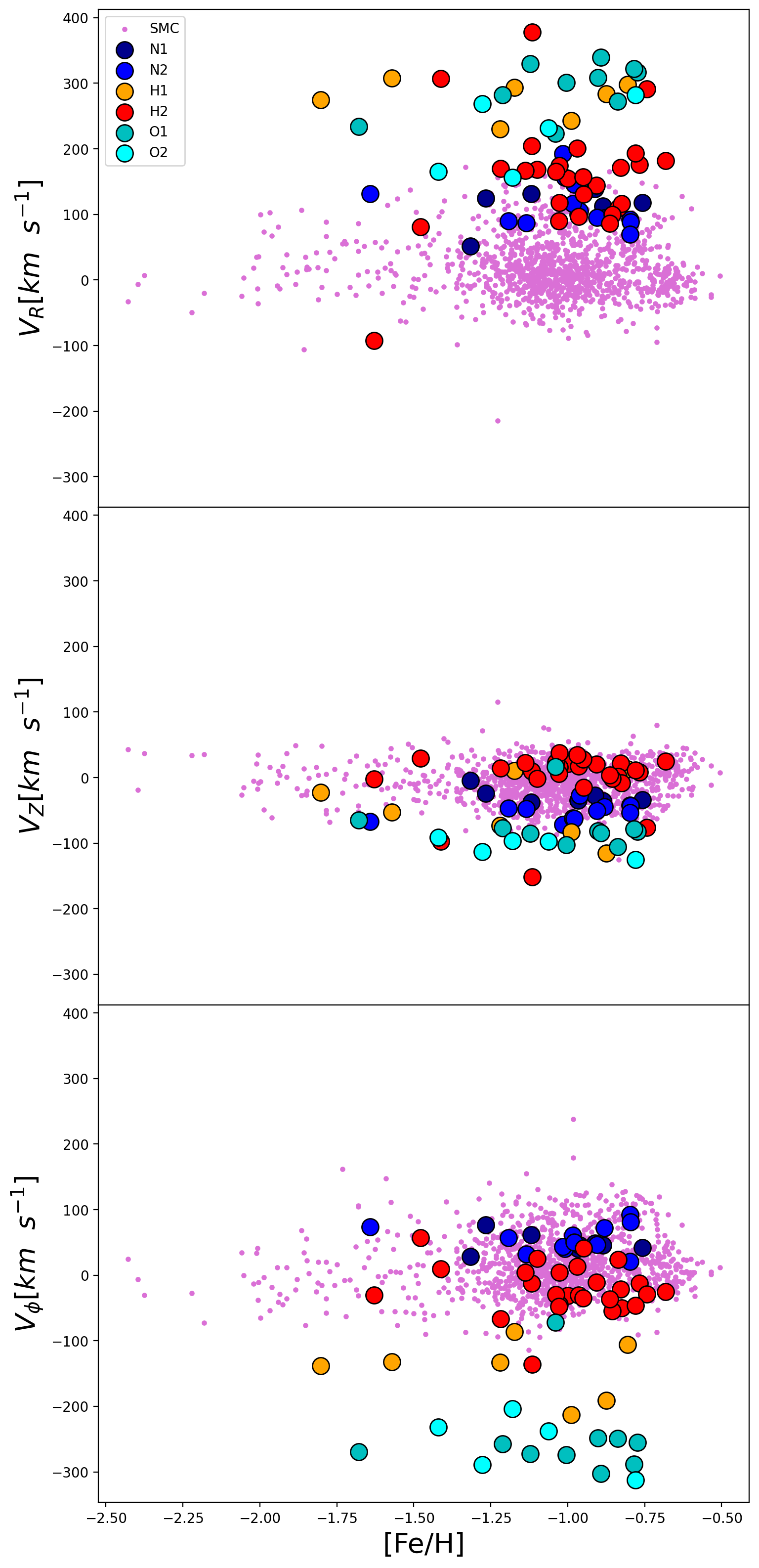} 
\caption{Vertical Velocity ($V_{z,{\rm SMC}}$), radial velocity ($V_{r,{\rm SMC}}$), and azimuthal velocity ($V_{\phi,{\rm SMC}}$) versus metallicity of the stars in the six targeted MC substructures along with stars targeted by APOGEE-2 in the central SMC.  
The velocities were calculated with respect  to the assumed SMC disk plane, as described in Section \ref{sec:data}, using an assumed distance of 60 kpc using the formalism and best-fit model parameters presented in \citet{Zivick2021}.}
\label{vel_zivick}
\end{figure}

\begin{figure*}
\centering
\includegraphics[width=0.98\hsize,angle=0]{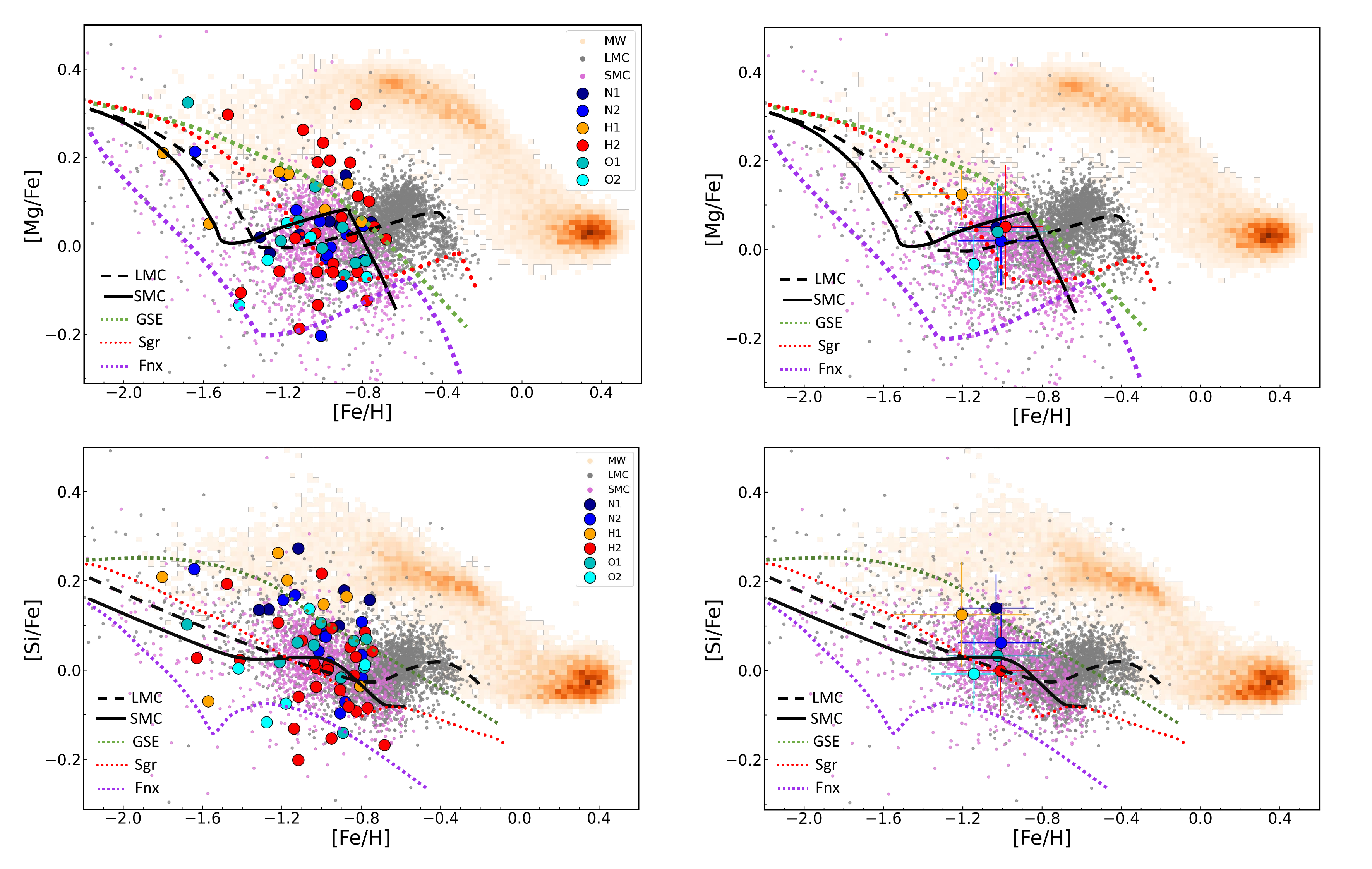}
\caption{A comparison of the distributions of [Mg/Fe] ({\it top panels}) and [Si/Fe] ({\it bottom panels}) versus [Fe/H] for our targeted MC substructure regions against those for large Milky Way satellites also using APOGEE data: the LMC, SMC, Sagittarius dSph, Fornax dSph, and the GSE.
The left panels show
individual stars for each substructure and the right panels show the mean and 
standard deviation for each substructure region. The over-plotted lines show the chemical evolution track for each  dwarf galaxy as determined by \citet{Hasselquist2021}.}
\label{trend_comp}
\end{figure*}

In addition to the RVs, thanks to the PMs delivered by Gaia, we can obtain 3-D kinematical information of the stars in the substructures by applying the models described in Section \ref{sec:data}. Figures \ref{vel_chen} and \ref{vel_zivick} show the 3-D velocities as a function of metallicities with respect to the LMC and SMC, respectively.  The mean 3-D velocity values for the six regions are summarized in Table \ref{tab:spacevelocities}.

\begin{table*}
\caption{Measured 3-D space velocities from the models relative to the LMC and SMC for our six regions}
\label{tab:spacevelocities} 
{
\begin{center}
\begin{tabular}{lcccccc}
\hline 
\hline
{\small{}Region} & 
$\langle V_{r,{\rm LMC}} \rangle$  & $\langle V_{\phi,{\rm LMC}} \rangle$ & $\langle V_{z,{\rm LMC}} \rangle$  & 
$\langle V_{r,{\rm SMC}} \rangle$  & $\langle V_{\phi,{\rm SMC}} \rangle$ & $\langle V_{z,{\rm SMC}} \rangle$
\tabularnewline
   & \kmse  & \kmse  & \kmse & \kmse  & \kmse  & \kmse   \tabularnewline

\hline 
LMC  &  4.4 & 61.3 & 9.7 &  &  &  \\
SMC  &  &  &  &  $-$10.0 & 8.5 & $-$15.2   \\

  N1  & $-$69.9 & 52.3 & 47.2 & 111.0& 65.3 & $-$28.2\\ 	

  N2  & $-$39.4 & 74.8 & 33.8 & 113.0 & 78.1 & $-$53.0 \\ 

 H1  & 2.0& 90.6& -16.3& 275.9 & $-$142.8 & $-$46.4 \\  	
 H2  & 13.4 & 53.4 & 42.4 & 157.0 & $-$21.2 &$-$ 0.2 \\ 
 O1  & $-$82.3 & 178.2 & $-$92.8 & 293.0 & $-$248.8 & $-$74.5 \\ 	
 O2  & $-$126.7 & 121.9 & 121.1 & 220.8 & $-$254.8 & $-$104.6 \\ 
\hline 

\end{tabular}
\end{center}
}
\end{table*}


C22's analysis of regions N1 and N2 using kinematics and MDFs implies that the APOGEE-targeted stars there have a strong relationship with the outer LMC disk (see also \citealt{Cullinane2020}) with the LMC having mean space velocity values of: $\langle V_r \rangle$=4.43 \kmse, $\langle V_{\phi} \rangle$=61.28 \kmse, and $\langle V_z \rangle$ = 9.70 \kms (see Figure \ref{vel_chen}). Indeed, comparing N1 and N2 with the 3-D LMC velocity distribution, we found that N1 and N2 show relatively good agreement for $V_\phi$ with a mean for N1 and N2 of $\sim$52.3 \kms and $\sim$74.8 \kmse, and especially for $V_z$ with a mean for N1 and N2 of $\sim$47.2 \kms and $\sim$33.8 \kmse, respectively. Additionally, these regions have relatively low velocity dispersions for the azimuthal and vertical velocity components of their stars with respect to the LMC (the dark blue symbols, with $\sigma$($V_{\phi}$)=11.00 \kms and $\sigma$($V_z$)=16.78 \kmse). On the other hand, comparing N1 and N2 with the 3-D SMC velocity distribution (see Figure \ref{vel_zivick}), with the SMC having mean space velocity values of: $\langle V_r \rangle$=$-$10.0 \kmse, $\langle V_{\phi} \rangle$=8.5 \kmse, and $\langle V_z \rangle$ = $-$15.2 \kmse, we find a more significant offset to these values for each velocity component in these regions. For $V_r$, the mean is $\sim$111.0 \kms and $\sim$113.4 \kms for N1 and N2 respectively, for $V_{\phi}$ the mean is $\sim$65.3 \kms and $\sim$78.1 \kmse. Finally, for $V_z$ the mean is $\sim$ $-$28.2 \kms and $\sim$ $-$53.0 \kmse, respectively. This indicates that N1 and N2 are kinematically more similar to the LMC than the SMC.
 
When comparing the 3-D velocity distribution of the LMC and SMC (Figures \ref{vel_chen} and \ref{vel_zivick}) with H1 and H2, we find a generally better agreement of these regions with the LMC, too.  In fact, we observe the most significant difference in the case of H1 and H2 when comparing the 3-D SMC velocity distribution for the case of $V_{r,{\rm SMC}}$, with a mean for H1 and H2 of $\sim$275.9 \kms and $\sim$157.0 \kmse, respectively.
Nevertheless, H2 appears to be also consistent with the SMC when looking at the $V_{\phi}$ and 
$V_{z,{\rm SMC}}$ velocities. 

O1 and O2 regions are the more complex regions, from a kinematical point of view.  The 3-D motion of their stars show the most significant discrepancy with both the LMC and SMC (see Figures \ref{vel_chen} and \ref{vel_zivick}). For O1, the mean velocities from the 3-D SMC model are $V_{r,{\rm SMC}}$=293.0 \kmse, $V_{\phi,{\rm SMC}}$=$-$248.8 \kmse, and $V_{z,{\rm SMC}}$=$-$74.50 \kmse,  while the mean velocities for the same components from the 3-D LMC model are $V_{r,{\rm LMC}}$=$-$82.31 \kmse, $V_{\phi,{\rm LMC}}$=178.22 \kmse, and $V_{z,{\rm LMC}}$=$-$92.8 \kmse. Similarly, for O2, the mean velocities from the 3-D SMC model are $V_{r,{\rm SMC}}$=220.78 \kmse, $V_{\phi,{\rm SMC}}$=$-$254.8 \kmse, and $V_{z,{\rm SMC}}$=$-$104.6 \kmse, while the mean velocities for the same components from the 3-D LMC model are $V_{r,{\rm LMC}}$=$-$126.67 \kmse, $V_{\phi,{\rm LMC}}$=121.9 \kmse, and $V_{z,{\rm LMC}}$=$-$121.1 \kmse  (See Table \ref{tab:spacevelocities}).
In C22, we found that these regions show a clear difference in their in-plane 
velocities, as obtained in C22 from the proper motion of the stars using the same kinematical model described in Section \ref{sec:data}, to those of
the LMC disk stars, with a difference of more than 100 \kmse.  This is also observed in the radial velocity ($V_r$), azimuthal velocity ($V_\phi$), and vertical velocity ($V_z$), obtained as described in Section \ref{sec:data} (see Figure \ref{vel_chen}). 
By comparing our results with numerical simulations of the MC interactions, we concluded in C22
that the stars in O1 and O2 are not simply an extension of the LMC disk, but most likely a combination of LMC and SMC tidally disrupted stars, although we could not
rule out other possible origins. We present here in this work also a 3-D kinematical SMC model. Even though there is discrepancy between the O1 and O2 values and those of the SMC from the 3-D model shown here in Figure \ref{vel_zivick}, it is worth noting that the SMC model, albeit having a tidal expansion, is an analytic model that does not completely capture the actual disruption and disturbances that the SMC is suffering. Thus, this might explain why the 3-D kinematical values for O1 and O2 differ from those of the SMC model.
Additionally, Figure 7 in C22 shows the results of two N-body simulations that model the past dynamical evolution and
interaction between the MCs performed by \citet{Besla2012}. The kinematics of O1 and O2 stars that we find here in Figure \ref{vel_zivick} may be accounted for as SMC tidal debris, according to Figure 7 in C22.

\section{Discussion}
\label{sec:discussion}

\begin{figure*}
\centering
\includegraphics[width=0.98\hsize,angle=0]{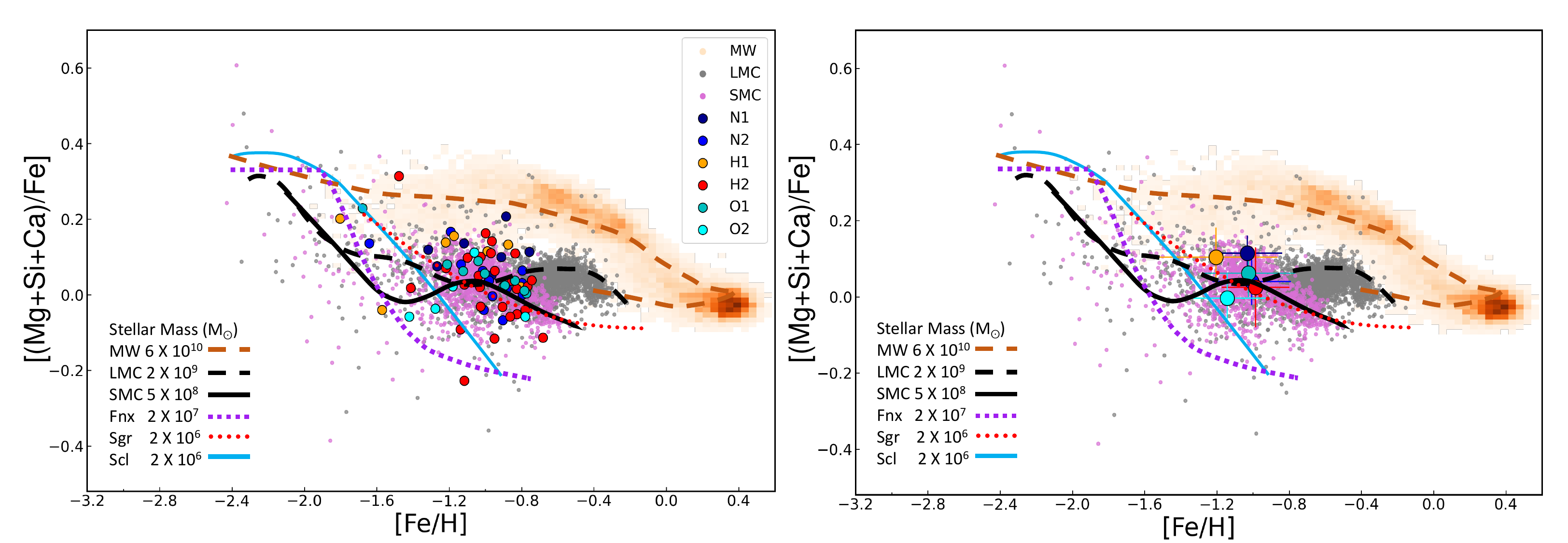}
\caption{The same as Figure \ref{trend_comp}, but for
[Mg+Si+Ca/Fe] versus [Fe/H].
Also included are the trendlines derived by \citet{Nidever2020} for the MW, LMC, SMC, Fornax (Fnx), Sagittarius (Sgr) and Sculptor (Scl).}
\label{trend_comp_Nid}
\end{figure*} 


Our chemical findings in the regions N1 and N2 (see Section \ref{subsec:n1n2}) indicate a slightly better agreement with the chemical patterns of the LMC. These findings are reinforced by the proper motion and radial velocity observed in those regions (see Figures \ref{Proper_Motion} and \ref{Met_Vel}), as well as with the 3-D motions (Figures \ref{vel_chen} and \ref{vel_zivick}), which show a better compatibility with the LMC. This is in line with the  complementary work by C22, who find no kinematical difference between these regions and the outer LMC. They are also in agreement with the work by \citet{Cullinane2020,Cullinane2021}, who studied several regions in the northeast outskirts of the LMC and analyzed the metallicity and kinematics along the northern arm.  They found a strong relationship between the properties in the northern arm and in the outer LMC, and suggest that over the last Gyrs the interaction between the LMC and MW produced the northern arm, especially considering the azimuthal velocity with positive out-of-plane values, which we also found in C22 and show here in Figure \ref{vel_chen}. It is noteworthy that they found, in the northern arm, that the metallicity decreases from [Fe/H]=$-$0.9 at 11 kpc to [Fe/H]=$-$1.2 dex at 22 kpc. This is in agreement with our findings for the LMC and for the northern substructures N1 and N2 (see Figure \ref{prof_LMC_SMC}). 


For regions H1 and H2 discussed in Section \ref{subsec:h1h2}, we find that H1 shows higher abundance values with respect to the LMC and SMC chemical pattern trendlines presented in Figures \ref{light} and \ref{iron}. However, region H1 shows a slightly better agreement with the LMC in $\alpha$-elements (Figure \ref{alpha}). This is supported by the proper motion and radial velocities observed in H1 (see Figures \ref{Proper_Motion} and \ref{Met_Vel}), as well as the 3-D derived velocities, which show a better match with the LMC. We caution that H1 has a  small sample size, with only seven stars in total and fewer for the Fe-peak elements.



The sample of H2 stars has chemical abundance patterns related to both the LMC and SMC, and it is the region with the largest chemical abundance dispersion (see Table \ref{tab:abundances}) for most of the elements, specifically in comparison with other regions, specifically for C, N, O, Al, Mg, Ca, V, Cr, Co and Ni. Also, we note a clear difference of the stars in H2 with respect to MW halo stars.  In C22, we found a strong kinematical relation between stars in H2 with the LMC disk. However, we find a good
resemblance to the SMC too from the 3-D velocities when comparing with the  model with respect to the SMC motion (see Figures \ref{vel_chen} and \ref{vel_zivick}).  Also, we note that the RVs (Figure \ref{Met_Vel}) and proper motions (Figure \ref{Proper_Motion}) could be related to either galaxy (LMC or SMC), with actually a larger number of H2 members in very good agreement with the SMC RVs.

\begin{figure*}
\centering
\includegraphics[width=0.85\hsize,angle=0]{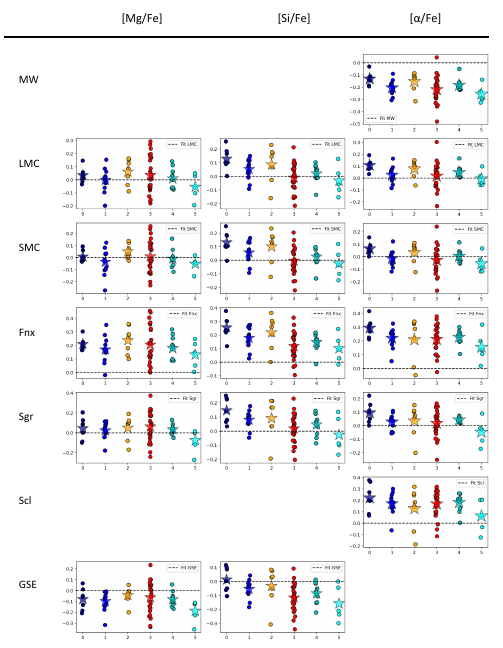}
\caption{Minimum orthogonal distance of each star from the six regions to the best fit for Mg-, Si-, and  $\alpha$-abundance versus metallicity for each galaxy in Figures \ref{trend_comp} and \ref{trend_comp_Nid} (MW, LMC, SMC, Fnx, Sgr, Scl and GSE).  The best fit was determined by \citet{Hasselquist2021} for Mg and Si, and by \citet{Nidever2020} for alpha-abundances. Each star from each region is represented by a filled circle and the mean for each region is represented by a filled star symbol (see Table \ref{tab:dist_nid_alpha}).}
\label{orthogonal_distance_obj}
\end{figure*} 




Regarding O1 and O2, we found that O1 has a better agreement in chemical abundance patterns with the SMC than with the LMC. O2 also appears to have similar chemical trends to the SMC, though there is a slight shift in the $\alpha$-abundance vs.\ metallicity towards lower metallicity values for O1. This may suggest that the stars in the O2 region belong to another substructure with a slightly different star formation onset time.  We note, however, that the analysis in O2 is done with only five stars and their $\alpha$-abundance values lie within the spread of the SMC values around the $\alpha$-abundance vs.\ metallicity trend (see Figure \ref{alpha}).

To gain more insight into the origins of the stars in the regions analyzed here, we show in Figures \ref{trend_comp} and \ref{trend_comp_Nid} the chemical evolution tracks of Mg, Si and [(Mg+Si+Ca)/Fe] presented by \citet{Hasselquist2021} and \citet{Nidever2020} for several galactic systems: the MW, LMC, SMC, Gaia Enceladus (GSE), Sagittarius (Sgr), Fornax (Fnx), and Sculptor (Scl), on top of our data. These studies also used APOGEE ASPCAP abundances so are very comparable to our study. Each evolutionary track describes the chemical evolution for the different systems presented by each author. The code used to perform the chemical evolution track by \citet{Hasselquist2021} and  \citet{Nidever2020} was flexCE \citep{Andrews2017}, which includes as parameters to perform the evolution tracks the initial gas mass, inflow rate, time dependence, and star formation efficiency.  Additionally, Figure \ref{orthogonal_distance_obj} presents the orthogonal distance of each star, colored according to the corresponding region, to the trendline of each system. The mean and standard deviation of this orthogonal distance are presented in Tables \ref{tab:dist_has_alpha} and \ref{tab:dist_nid_alpha}. This orthogonal distance  helps  us to analyze the agreement of each star, and thus each region, to each of the systems presented in these figures.

N1 shows a good agreement with the evolution track of Mg and [(Mg+Si+Ca)/Fe]  for the LMC and SMC (Figures \ref{trend_comp} and \ref{trend_comp_Nid}). If we take into account the mean orthogonal distance to the evolution tracks amongst all of the systems analyzed and presented in Tables \ref{tab:dist_has_alpha} and \ref{tab:dist_nid_alpha}, N1 shows the best agreement with the LMC for the $\alpha$-elements, and N2 presents a similar behavior. 

Comparing the chemical evolution tracks with H1, we find a reasonably good agreement of this region with both the LMC and SMC. However, in Tables  \ref{tab:dist_has_alpha} and \ref{tab:dist_nid_alpha}, we show that  H1 is closer  to the  LMC model for the case of Mg, Si and Alpha ([(Mg+Si+Ca)/Fe]). For H2, the minimum distance agrees with the LMC and SMC with a slight difference between them.

O1 shows similar behavior to the other regions. The minimum orthogonal distance is to the LMC model in the case of Mg, Si and $\alpha$-abundance. However, note that the most metal-poor star in region O1 (at [Fe/H] $\sim$ $-$1.7), is also the most Mg-rich and the most Si-rich among the O1 stars, in complete agreement with the chemical evolution track of the SMC presented by \citet{Nidever2020} and \citet{Hasselquist2021}.  The bump in the SMC chemical evolution track with a peak in metallicity at [Fe/H] = $-$0.9 is an indication of the highest SFR rate and that this major SMC burst happened $\sim$4 Gyrs earlier than the burst in the LMC, according to Hasselquist et al.\ chemical track models.
This bursting was dominated by SNII, therefore, we observe an enhancement in material associated with SNII. After the peak at [Fe/H] = $-$0.9, we observe a depletion in these materials due to low contribution from SNII and the onset of SNIa, and this behavior is observed in the most metal-poor star in O1 for Mg and Si.  Although one star in Ca shows a significant discrepancy, this star is the most Ca-rich with a metallicity about [Fe/H]=$-$0.9 dex, but this star generally exhibits good agreement with both MCs in the other $\alpha$-elements.

O2 exhibits the most peculiar behavior compared to both models. We found O2 showed the minimum distance in the case of Si with SMC and almost identical to that of the Sgr model, but with a larger standard deviation in this case. For Mg, we observe a similar behavior, and for the $\alpha$-model we observe that the minimum distance is with Sgr, but again, with only a slight difference and larger standard deviation than that of the SMC model chemical track. For O2, the increase seen in Mg and Ca starting from the most metal-poor star, at [Fe/H]=$-$1.4 is, again, an indication of the enrichment in $\alpha$-elements produced mainly by SNII. After that we observe one star at [Fe/H]=$-$1.1, which is the most $\alpha$-rich star in O2 and at which we see the increased trend in $\alpha$-abundance ending (see Figures \ref{trend_comp} and \ref{trend_comp_Nid}). At higher metallicities, there is only one star in O2, the most metal rich-one, that shows a lower $\alpha$-abundance. This could be an indication of enrichment by SNII in the range of metallicity between [Fe/H]=$-$1.4 and $-$1.1 at which point the contribution from SNII starts to decrease. This trend is slightly shifted from the $\alpha$-abundance vs. metallicity trend found in the SMC (see Figure 20 in \citealt{Nidever2020} and \citealt{Hasselquist2021}) and even more shifted than that in the LMC, such that the contribution of SNIa starts at progressively lower metallicities of [Fe/H]=$-$0.4, $-$0.9, $-$1.1 in  the LMC, SMC and O2, respectively (Figure \ref{alpha}).


\begin{table*}
  \centering
  \caption{Summary information for the six regions}
  \label{tab:table_fields}
  \begin{tabular}{l|cc|cc|cc|cc|c}
    \hline
    \textbf{Region} & \multicolumn{2}{|c|}{MDF} & \multicolumn{2}{|c|}{Radial Gradients} & \multicolumn{2}{|c|}{Kinematics} & \multicolumn{2}{|c|}{Chemistry} & \textbf{Origin} \\
    & \textbf{L} & \textbf{S} & \textbf{L} & \textbf{S} & \textbf{L} & \textbf{S} & \textbf{L} & \textbf{S} & \\
    \hline
    N1 &  \checkmark & \checkmark   &  \checkmark & $\times$  &  \checkmark & $\times$   &  \checkmark & \checkmark  & LMC \\
    N2 &  \checkmark & \checkmark   &  \checkmark & $\times$  &  \checkmark & $\times$   &  \checkmark & \checkmark  & LMC \\
    H1 &  \checkmark & \checkmark   &  \checkmark & $\times$  &  \checkmark & $\times$   &  \checkmark & \checkmark  & LMC \\
    H2 &  \checkmark & \checkmark   &  \checkmark & \checkmark  &  \checkmark & $\sim$   &  \checkmark & \checkmark+  & SMC+LMC \\
    O1 & $\times$ & \checkmark   &  \checkmark & $\times$  &  $\times$ & $\sim$   &  \checkmark & \checkmark+  & SMC+LMC \\
    O2 & $\times$ & \checkmark   &  $\times$ & \checkmark  &  $\times$ & $\sim$   &  \checkmark & \checkmark+& SMC \\
    \hline
  \end{tabular}
  \tablefoot{The checks and cross marks indicate whether the stars are compatible with the LMC (L) or SMC (S) in this particular attribute. The plus sign next to a check indicates better agreement. $\sim$ indicates that it is marginally compatible.}
\end{table*}

Table \ref{tab:table_fields} summarizes the potential origins of our regions based on combining the results of four different diagnostics: MDF, radial gradients, 3-D kinematic modeling, and chemical abundances pattern. It is evident that all of the regions have reasonable connections to both the SMC and LMC, but in particular half of the regions are clearly of LMC origin, namely N1, N2, and H1. These are the two northern regions -- along the northern ``stream''-like feature -- and the southern region closest to the LMC.  Regions H2 and O1 show compatibility with both the LMC and SMC and thus are likely of tidal origin from both galaxies. The O2 field, located in the southern region closest to the SMC, is less compatible with the LMC, not only from the radial gradients, MDF and the 3-D kinematic model, but also from the chemical abundance pattern analyzed. We conclude that this region contains SMC tidally perturbed stars.

\section{Summary and Conclusions}
\label{sec:summary}

In this paper, we study the chemistry of six regions located in substructures in the periphery of the Magellanic Clouds, with the main goal of trying to understand the origin of their stars. These regions were previously analyzed in a companion paper from a 3-D kinematical point of view \citet[][hereafter C22]{Cheng2022}, but only from an LMC kinematical reference frame. Our analysis now focuses on the detailed chemical patterns exhibited by different elements in comparison with both of the Magellanic Clouds and we also add a 3-D kinematical model based on the SMC reference frame. We have used data from the near-IR APOGEE-2 spectrograph, which allowed us to collect the chemical abundances of 13 different elements, including light, $\alpha$ and Fe-peak, with a signal-to-noise from  35 to 192 for a total of 69 red giants in these regions. These data correspond to a sub-sample of the data presented in the C22 study, where only the high S/N stars are now considered.

Our detailed chemical pattern analysis in conjunction with 3-D kinematical information suggests the following about each region:


\begin{itemize}

\item N1 and N2, the two regions along the northern LMC ``stream''-like feature, with 7 and 13 members, respectively, show the strongest relationship with the LMC.  We find a good agreement between the chemical patterns of N1 and N2 and those of the LMC for light, $\alpha$ and Fe-peak elements which confirm from a chemical point of view that N1 and N2 stars are thus perturbed outer LMC disk stars. 



\item H1 and H2, with 7 and 27 members, respectively, belong to the southern periphery of the LMC.  We find that H1 is the most metal-poor of the six regions, albeit with a large scatter. Overall, we have found that the mean abundances of $\alpha$, light, and Fe-peak elements exhibit a significant difference, being enhanced  compared to the trend observed in the MCs.

From the kinematical point of view, we note that H1 shows good agreement with the LMC. On the other hand, H2 shows 3-D motions associated with both LMC and SMC. We conclude that H2 is likely populated with stars from both the LMC and SMC. 

\item O1 and O2 have 10 and five star members, respectively. Their chemical patterns of $\alpha$, light and Fe-peak elements  are in broad agreement with both MCs, however they are more consistent with the SMC chemical evolution track than with the LMC. This is also the case when comparing the MDFs and the radial metallicity profiles. These two regions are more complex regions, in terms of kinematical behavior, showing clear differences in their 3-D modeled velocities from both the LMC and SMC reference frames. However, numerical models show that the kinematics of O1 and O2 may be accounted for as SMC tidal debris. This, together with the better chemical pattern agreement with the SMC, lead us to conclude that the stars in these regions, particularly in O2, are of an SMC origin.


Additionally, in this work:

\item We present for the first time a metallicity and an alpha-abundance radial profile for the LMC and SMC galaxies, extending to distances of up to 20\dgr and 10\dgr, respectively. The slopes of the metallicity gradients are $-$0.03 and $-$0.04 dex deg$^{-1}$, respectively. We also find positive alpha abundance gradients in both galaxies, with slopes of 0.04 and 0.01 dex deg$^{-1}$, respectively(see Figure \ref{prof_LMC_SMC}).

Our findings indicate that regions N1 and N2 are clearly LMC stars, confirming the kinematical analysis by C22, that were removed from the outer disk possibly due to the interactions of the LMC with the MW in its first pericenter passage (see also \citealt{Cullinane2021}). The southern region H1 is also likely of LMC origin, and H2 is a mix of LMC and SMC stars, with a preference for the region H2 being dominated by SMC stars.  It is also probable that H1 has some MW halo stars contaminating our sample. The regions O1 and O2 are populated by a mix of LMC and SMC stars that were likely tidally disrupted due to the interaction of both MCs. The O2 region, in particular, shows a chemical abundance pattern very similar to that of the SMC and velocities more similar to SMC debris, and thus, we conclude that these stars are of SMC origin. Finally, this study highlights the importance of having chemical abundances in addition to kinematics to help confirm the nature of the stars in the outskirts of the LMC and provide evidence that can be used to better constrain the interaction history of the MCs as well as to improve our knowledge of their orbital history.

\end{itemize}

\begin{acknowledgements}
      We thank the anonymous reviewer for their helpful comments, which improved the quality of this paper. C.M. thanks the support provided by FONDECYT Postdoctorado No.3210144. A.M. gratefully acknowledges support by the ANID BASAL project FB210003,  FONDECYT Regular grant 1212046, and funding from the Max Planck Society through a “PartnerGroup” grant. 
D.L.N. acknowledges support for this research from National Science Foundation (NSF) grant AST-1908331, while X.C, S.R.M., and A.A. acknowledge NSF grant AST-1909497.

This work has made use of data from the European Space Agency (ESA) mission {\it Gaia} (\url{https://www.cosmos.esa.int/gaia}), processed by the {\it Gaia} Data Processing and Analysis Consortium (DPAC, \url{https://www.cosmos.esa.int/web/gaia/dpac/consortium}). Funding for the DPAC has been provided by national institutions, in particular the institutions participating in the {\it Gaia} Multilateral Agreement.

Funding for the Sloan Digital Sky Survey IV has been provided by the Alfred P. Sloan Foundation, the U.S. Department of Energy Office of Science, and the Participating Institutions. 

SDSS-IV acknowledges support and resources from the Center for High Performance Computing  at the University of Utah. The SDSS website is www.sdss.org.

SDSS-IV is managed by the Astrophysical Research Consortium for the Participating Institutions of the SDSS Collaboration including the Brazilian Participation Group, the Carnegie Institution for Science, Carnegie Mellon University, Center for Astrophysics | Harvard \& Smithsonian, the Chilean Participation Group, the French Participation Group, Instituto de Astrof\'isica de Canarias, The Johns Hopkins University, Kavli Institute for the Physics and Mathematics of the Universe (IPMU) / University of Tokyo, the Korean Participation Group, Lawrence Berkeley National Laboratory, Leibniz Institut f\"ur Astrophysik Potsdam (AIP),  Max-Planck-Institut f\"ur Astronomie (MPIA Heidelberg), Max-Planck-Institut f\"ur Astrophysik (MPA Garching), Max-Planck-Institut f\"ur Extraterrestrische Physik (MPE), National Astronomical Observatories of China, New Mexico State University, New York University, University of Notre Dame, Observat\'ario Nacional / MCTI, The Ohio State University, Pennsylvania State University, Shanghai Astronomical Observatory, United Kingdom Participation Group, Universidad Nacional Aut\'onoma de M\'exico, University of Arizona, University of Colorado Boulder, University of Oxford, University of Portsmouth, University of Utah, University of Virginia, University of Washington, University of Wisconsin, Vanderbilt University, and Yale University.
D.G. gratefully acknowledges support from the ANID BASAL project ACE210002.
D.G. also acknowledges financial support from the Direcci\'on de Investigaci\'on y Desarrollo de
la Universidad de La Serena through the Programa de Incentivo a la Investigaci\'on de
Acad\'emicos (PIA-DIDULS).
R. R. M. gratefully acknowledges support by the ANID BASAL project FB210003.
J.G.F-T gratefully acknowledges the grant support provided by Proyecto Fondecyt Iniciaci\'on No. 11220340, and also from ANID Concurso de Fomento a la Vinculaci\'on Internacional para Instituciones de Investigaci\'on Regionales (Modalidad corta duraci\'on) Proyecto No. FOVI210020, and from the Joint Committee ESO-Government of Chile 2021 (ORP 023/2021), and from Becas Santander Movilidad Internacional Profesores 2022, Banco Santander Chile. 
\end{acknowledgements}

%
%

\bibliographystyle{aa}

\bibliography{biblio.bib}{}
  

\end{document}